# Computation of a Theoretical Membrane Phase Diagram, and the Role of Phase in Lipid Raft-Mediated Protein Organization

Eshan D. Mitra[†1], Samuel C. Whitehead[‡], David Holowka[†], Barbara Baird[†*], James P. Sethna[‡*]

[†]Department of Chemistry and Chemical Biology, Cornell University, 122 Baker Laboratory, Ithaca, NY, 14853
[‡]Department of Physics, Cornell University, 109 Clark Hall, Ithaca, NY, 14853

[1]Current address: Los Alamos National Laboratory, P.O. Box 1663, Los Alamos, NM 87545

*Corresponding authors:
B. Baird, bab13@cornell.edu
J. P. Sethna, sethna@cornell.edu

## Abstract

Lipid phase heterogeneity in the plasma membrane is thought to be crucial for many aspects of cell signaling, but the physical basis of participating membrane domains such as "lipid rafts" remains controversial. Here we consider a lattice model yielding a phase diagram that includes several states proposed to be relevant for the cell membrane, including microemulsion – which can be related to membrane curvature – and Ising critical behavior. Using a neural network-based machine learning approach, we compute the full phase diagram of this lattice model. We analyze selected regions of this phase diagram in the context of a signaling initiation event in mast cells: recruitment of the membrane-anchored tyrosine kinase Lyn to a cluster of transmembrane of IgE-FcεRI receptors. We find that model membrane systems in microemulsion and Ising critical states can mediate roughly equal levels of kinase recruitment (binding energy ~ -0.6 $k_BT$), whereas a membrane near a tricritical point can mediate much stronger kinase recruitment (-1.7 $k_BT$). By comparing several models for lipid heterogeneity within a single theoretical framework, this work points to testable differences between existing models. We also suggest the tricritical point as a new possibility for the basis of membrane domains that facilitate preferential partitioning of signaling components.

**Introduction**

The lateral organization of cell plasma membranes, which contributes crucially to their functions, is regulated by membrane proteins and lipids as well as by attachment to the cytoskeleton and by communication with membrane trafficking and other cellular processes. A primary component of membrane organization appears to be the collective properties of the lipid populations, and this has been examined experimentally and theoretically, as described in numerous recent reviews (see [1,2] and reviews cited therein and elsewhere in this paper). Whereas the diameter of a constituent lipid is about 1 nm, the bulk of experimental evidence suggests that mammalian plasma membranes contain phase-based domains on the order of 10-200 nm in length.[3–6] This heterogeneity has been related to studies of simpler model membranes composed of a high melting point ($T_m$) lipid, a low $T_m$ lipid and cholesterol, considered to serve to as an approximation of plasma membrane lipids.[7] Varying relative amounts of these three types of lipids has yielded phase diagrams showing regions of separation between phases characterized as liquid ordered (Lo, more high-$T_m$ lipid and more cholesterol) and liquid disordered (Ld, more low-$T_m$ lipid).[8–10]

Studies on giant plasma membrane vesicles (GPMVs), which are isolated from cells, exhibit micrometer-scale regions of Lo-like and Ld-like phase character.[11] Similar fluorescence microscopy studies on intact cells under physiological conditions do not detect Lo/Ld separation above the diffraction limit, possibly due in part to their dispersal by cytoskeletal attachment in cells.[12] However, electron spin resonance (ESR) studies on intact cells provide evidence for coexisting Lo and Ld domains.[13] In cell plasma membranes these nanometer-scale phase-like domains are thought to be coalesced or stabilized as a result of an external stimulus (e.g. antigen cross-linking of immune receptors), and to play an essential role in stimulated cell signaling, by facilitating colocalization of membrane proteins that partition into the same Lo-like domain, and separating them from those that partition into Ld-like domains.[6,14] We are particularly interested in cases where induced interactions between multiple Lo-preferring components stabilize these domains, thereby recruiting other Lo-preferring components. Such lipid-mediated segregation has been implicated in many mechanisms of membrane protein signaling, including immune receptors,[15,16] G-protein coupled receptors,[17] the oncogenic GTPase Ras,[18] and others. A generic term that has emerged for plasma membrane domains of Lo-like character is "lipid rafts," and, although the size, dynamics, and other features of these structures in functional cells surely vary compared to those in model membranes, the lipid phase properties are expected to be similar.

### *Theories of raft formation*

Despite the centrality of lipid-based membranes to cell biology, there remains no consensus on the physical basis of lipid domains. As described above, formation of lipid rafts has been tied to the observation of phase separation in model plasma membranes, including giant unilamellar vesicles (GUVs)[8–10,19,20] and GPMVs[11,21]. In addition to the simplest forms of two-phase coexistence, these systems exhibit a rich variety of phase behavior, including, microemulsions[a],[19,22] lamellar phases (also called modulated phases),[23] and critical phenomena.[21] Moreover, despite recent advances in experimental techniques (for recent reviews, see [16,24]), lipid rafts in cell plasma membranes remain a difficult system to investigate – the dynamics and complexity of real cell systems notwithstanding, the 10-200 nm dimension of rafts[4] prevents direct observation via conventional light microscopy. Thus, the goal for a theoretical consideration of lipid raft physics should provide comparisons and hypotheses that are amenable to testing with the currently available tools.

Towards this end, various theoretical models have been proposed to describe raft-like phenomena. However, due to the lack of direct experimental data on lipid rafts, the set of theories that are consistent with observation is relatively unconstrained – models that disagree on the fundamental physics of raft formation can give qualitatively similar results that agree with extant experimental work.[1] One theoretical viewpoint is that lipid rafts are mediated by membrane curvature,[25–27] which makes the interface between immiscible membrane domains more energetically stable. It has also been proposed that a surfactant species could provide a similar interface between domains.[28] Both of these viewpoints suggest that rafts exist as part of a microemulsion phase, in which nanoscopic domains of a characteristic size are stabilized due to the curvature or surfactant. An alternate hypothesis

---

a   Some groups describe the presence of "nanodomains,"[19] a state of *two-phase coexistence* consisting of nanoscopic domains of a characteristic size, rather than a microemulsion, which is defined as a *one-phase* state with domains of a characteristic size.

The difference in terminology arises from a difference in the definition of the location of the phase boundary. Theoretical physicists commonly define a phase based on the average of some order parameter, which is calculated over a long length scale. If this length scale is larger than the characteristic domain size, then the domains are averaged out in this calculation, leading to the conclusion that the system consists of a single phase, and the designation of "microemulsion". However, some experimental groups define a system to be in two-phase coexistence whenever an experimental technique (e.g. FRET, which has a detection length scale of ~2-8 nm) detects the presence of two components. [22] Analysis of the same "microemulsion" system with small characteristic domains would indeed give detection of two distinct components, leading to the conclusion of two-phase coexistence, and the label of "nanodomains."

In this study, we use the term "microemulsion," but note that the same area of the phase diagram could be deemed "nanodomains" if one adopted an empirical definition of two-phase coexistence such as is used in [22].

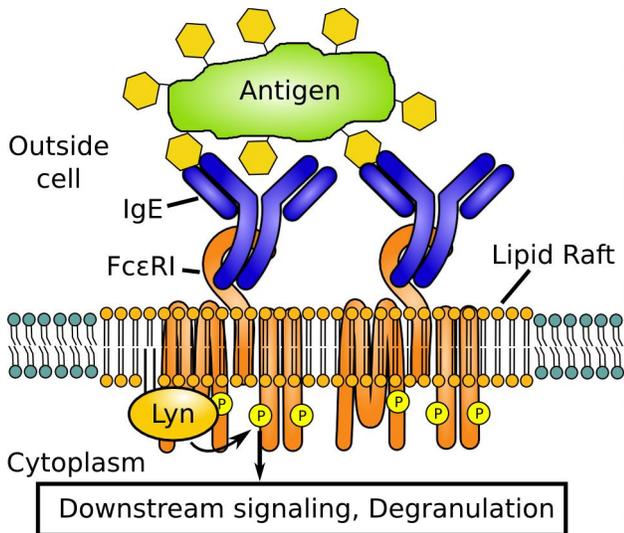

**Figure 1**. Signal initiation by IgE-FcεRI. IgE-FcεRI are cross-linked by an external antigen. The resulting cluster of receptors stabilizes a lipid raft that enables the recruitment of Lyn. Lyn performs the initial phosphorylation steps that transmit the signal to more downstream signaling partners.

suggests that rafts are formed from critical fluctuations in membrane composition, a result of proximity of the membrane to a 2D Ising critical point.[21,29] Experimental studies have provided support for both an Ising critical point[21] and interactions consistent with a microemulsion.[22,23]

We define lipid rafts as nanoscale domains concentrated with Lo-preferring components, and as described above, these domains can serve to colocalize membrane proteins that partition similarly.

To analyze the functional consequence of rafts in depth, we focus on the tractable example of transmembrane signaling mediated by the IgE-FcεRI receptor in mast cells. Physiologically, this stimulated transmembrane coupling activates cellular signaling pathways involved in allergic immune responses (reviewed in [30–32]). The mast cell is stimulated when specific multivalent ligands (antigen) physically cross-links several IgE-FcεRI receptors together in a cluster. This clustering causes recruitment of the kinase Lyn, which is anchored to the inner leaflet of the plasma membrane and when recruited, phosphorylates the receptor, thereby activating downstream signaling events (Figure 1). This kinase recruitment is thought to be raft-mediated: both the cross-linked receptors and the kinase preferentially partition into Lo-like membrane domains, facilitating their coupling on the plasma membrane.[15,33,34] The mast cell system serves as an example of a more general paradigm in cell biology, in which the orchestrated co-clustering of membrane proteins due to an external stimulus leads to initiation of transmembrane signaling.[18]

## Model

In this work, we address some of the ambiguities in the physics of lipid raft formation by a comparative approach. We employ a lattice model originally described by Gompper and Schick,[35,36] which can be used for simultaneous evaluation of both microemulsions and critical phenomena. In addition, this model captures features such as a lamellar (modulated) phase and two-phase coexistence observed in other membrane studies. Moreover, the model exhibits a tricritical point – defined as the termination of a three-phase coexistence regime in a phase diagram – which we suggest has interesting implications for stimulated cell signaling.

The model consists of a two-dimensional square lattice with the Hamiltonian (Eq. 1 with variables defined below)

$$H = \sum_i \left(H\sigma_i + \Delta\sigma_i^2\right) + \sum_{i,j}\left(-J\sigma_i\sigma_j - K\sigma_i^2\sigma_j^2\right) + \sum_{i,j,k} L\sigma_i\left(1-\sigma_j^2\right)\sigma_k \qquad (1)$$

Each site on the 2D lattice can take a spin value $\sigma$ of -1 (black pixel), 0 (gray pixel), or 1 (white pixel). Black and white pixels represent membrane components favoring Ld and Lo domains, respectively. Gray pixels represent a surfactant when surfacant strength $L$ is greater than 0, or simply a molecule with neutral domain preference when $L = 0$. The summation over $i$ is over all sites in the lattice; $i,j$ is over all nearest neighbors; $i,j,k$ is over all groups of three adjacent pixels in a straight horizontal or vertical line. We equate one lattice unit to a length of 1 nm, the approximate diameter occupied by one membrane lipid molecule.

Each of the five parameters of this model – $H$, $\Delta$, $J$, $K$, and $L$ – has units of energy. We consider only non-negative values for $J$, $K$, and $L$, while $H$ and $\Delta$ can take any value. The external fields $H$ and $\Delta$ control the composition of the lattice. $H$ controls the relative abundance of $\sigma$=-1 (black pixels) and $\sigma$=1 (white pixels), while larger $\Delta$ increases the concentration of $\sigma$=0 (gray pixels). We consider this model in the grand canonical ensemble: our simulation box represents one section of the membrane, so it makes sense that the number of each type of molecule can fluctuate, analogous to molecules diffusing in and out of the box. The coupling $J$ between adjacent pixels represents the usual Ising model coupling, which, for a membrane model, is the preference for molecules that prefer Lo domains to be adjacent to other molecules that prefer Lo domains (and similarly for Ld-preferring molecules). $J$ can also be thought of as equal to the line tension between black and white pixels times a distance of 1 lattice unit (1 nm). $K$ is a two-pixel interaction that gives a favorable energy to adjacent non-gray pixels. For a particular concentration of gray pixels, a higher value of $K$ makes it more favorable to have those gray pixels adjacent to each other. $L$ controls the strength of gray pixels as a surfactant; this term contributes a nonzero value only when a gray pixel ($\sigma$=0) sits between two non-gray pixels ($\sigma$=±1), and is favorable when the two non-gray pixels have different signs. Thus, increasing $L > 0$ makes it more favorable for gray to sit between black and white.

In our implementation, rather than choosing a value for $J$, we choose a value for temperature $T$ in units of $J/k_B$, and $J$ is set accordingly. The other parameters $H$, $\Delta$, $K$ and $L$ are chosen in units of $J$. Boltzmann's constant $k_B$ is set to unity.

Note that when $L=0$ Eq. 1 reduces to the Blume-Emery-Griffiths model.[37] With $K$ also set to 0, Eq. 1 becomes the Blume-Capel model.[38,39] With $\Delta = -\infty$, corresponding to no gray pixels present, Eq. 1 reduces to the Ising model.

### *Phases in the Lattice Model*

When the lattice model of Gompper and Schick was initially described, it was possible to extract some key features of the phase diagram, most notably the location of the critical line, by finite size scaling.[35] With the great increase in the power of computational resources since that time, it has become possible for us to address the model more globally by simulation.

We further take advantage of neural networks, which have become a powerful machine learning technique, leading to the development of computational tools to address challenging problems such as image recognition.[40] In image recognition, a neural network is trained to read the pixel values of an image, and output a label corresponding to what the image shows, such as distinguishing between a cat and a dog. Similarly, neural networks have been trained on simulated snapshots of phase models in physics, to output a label corresponding to which phase the snapshot represents. When this phase classification is performed for snapshots at a large number of model parameter sets, one obtains the phase diagram of the model.[41,42] With this methodology, we label a region of parameter space as a distinct "phase" if the neural network is able to distinguish simulation snapshots in that region from snapshots representing other phases. This definition is not always equivalent to a thermodynamic definition of a phase (i.e. based on the value of order parameters), but rather puts a greater emphasis on visually identifiable, qualitative differences in system properties.

Based on our neural network analysis, we describe eight phases (distinguishable qualitative behaviors specified below) that the model (Eq. 1) produces. We name these as follows: Within the *fluid* phase, all three components are well-mixed, with only short-range interactions between them. The *black* phase and *white* phase consist of nearly all black pixels and white pixels, respectively. When $H=0$, the Hamiltonian (Eq. 1) is symmetric with respect to exchanging black and white, and so these phases are seen in a state of two-phase coexistence. The *gray* phase consists of nearly all gray pixels. The *microemulsion* "phase" consists of black and white domains stabilized by a boundary of surfactant. The *critical* "phase" consists of fluctuating black and white domains, resulting from close proximity above a critical phase transition. Note that the microemulsion and critical "phases" are not thermodynamically distinct from one another or from the fluid phase.[b] [7,43] However, because we are interested in the qualitative nature of domains that could be relevant for membranes, we choose to consider them separately. The *lamellar* phase, is similar to a microemulsion "phase" in that the surfactant separates the black and white domains, but instead of enclosed, roughly round domains, the two domains exist as long stripes. Finally, the *crystal* phase includes the behavior in which rectangular domains of black and white exist, separated by a meshwork of surfactant.

### *Applying the model to cell signaling*

Here, we apply the methodology of neural networks to the Gompper and Schick lattice model, with the ultimate goal of understanding how different qualitative phase behaviors in membranes compare in their capacities to mediate cell signaling through membrane receptors (Figures 1 and 2c). The neural-network-derived phase diagram labels regions of parameter space according to their distinctive behaviors, as described in the previous section. We use this diagram to focus on sections of parameter space that are proposed to be relevant for plasma membrane heterogeneity, in particular the microemulsion and critical "phases". At these interesting points, we perform Monte Carlo simulations to calculate the energy associated with recruitment of an inner-membrane-anchored kinase (Lo-preferring) into a transmembrane receptor cluster (also Lo-preferring), as in the mast cell signaling system. Note that these recruitment energies—in contrast to binding energies associated with chemical bonds—are associated with long range forces: Proteins are recruited into an energetically favorable region, without orienting and binding directly to specific sites on proteins that stabilized the energetically favorable region. Also the energies we calculate are non-specific – Lo and Ld preferring proteins will share the same interactions as a group, and their structure details would only determine the degree of preference. Thus these long range forces allow nonspecific interactions that are restricted only in terms of the components' phase preference, as for co-localization in lipid rafts.

Monte Carlo methods allow us to explore the protein energetics semi-quantitatively throughout the phase diagram. Moreover, the recruitment energies that we calculate agree with exact conformal field theory results near the Ising critical point,[44] and hence should quantitatively describe experimental systems near critical points. And while our simulations focus on a simplified model of clustered receptors, near critical points our results are

---

b   Those who define microemulsion-like states as two-phase nanodomains (see footnote a), would instead say that the microemulsion and critical "phases" are part of the two-phase coexistence between the black and white phases.

universal, and are thus generalizable to a broad range of phenomena associated with membrane heterogeneity. In total, this method of recruitment energy calculation allows us to evaluate how the qualitative behavior of the plasma membrane relates to its capacity to form lipid rafts that can be stabilized (e.g., by clustered receptors) to mediate biologically relevant signaling.

The neural network approach is uniquely suited for this goal, offering a number of advantages over more traditional analysis approaches. First, it is capable of exploring large areas of parameter space at low computational cost. Second, it is able to detect qualitative changes in model behavior, such as microemulsions, even if those changes do not correspond to a true thermodynamic phase transition. These qualitative differences have important consequences for cell signaling that is facilitated by membrane organization.

**Methods**

*Monte Carlo Simulations*

Snapshots of the lattice model[35] were generated by the Metropolis algorithm. The length of the simulation was counted in sweeps, where, in each sweep, each lattice site has on average 100 opportunities to be flipped (total of 90000 individual proposed moves for a 30x30 lattice,[c] Figure 2a). Each proposed move consisted of randomly choosing a lattice site and a target value (one of {-1, 0, 1} that was not the current value at the site). The move was performed with probability min(1, $e^{-\Delta U/T}$) where $\Delta U$ is the change in the Hamiltonian energy (Eq. 1) resulting from the move.

To generate a single independent snapshot, the lattice was randomized, then 100 sweeps were run to equilibrate, and the final result was saved. To generate correlated snapshots, additional sweeps were run after equilibration, and a sample was saved after each sweep. Such snapshots are correlated because a single sweep is not enough to fully reequilibrate the lattice.

*Neural Network Training*

We chose the cross-section $H/J$=0, $K/J$=2, $L/J$=3 (see Hamiltonian, Eq. 1) for training

---

c  The lattices we use for mapping phase diagrams are small; the size was chosen to capture the correlations on length scales of interest to protein aggregation, and for convenient training of the network. Phases without structure on long length scales should be well described by our small simulations; we would expect shifts in boundaries of microemulsion phases, for example, only when the modulation approaches 30 pixels. Near critical points all length scales are important for the physics, but we show that the phase boundaries converge fairly rapidly. The shift in the effective critical temperature in a system of size $L$ goes as $L^{1/\nu}$, so for the Ising critical point with $\nu$=1 we expect 3% shifts in phase boundaries for a 30x30 system (beyond the precision of our methods), and near the tricritical point with $\nu$=5/9 we find even smaller shifts.

because this is close to the cross-section described by Gompper and Schick[35] as containing examples of all major phases of the model. Generation of the neural network training data was an iterative and somewhat heuristic process. We started by sparsely sampling a large region of ($\Delta$,$T$) space in the $H/J$=0, $K/J$=2, $L/J$=3 plane and labeling phases manually, to get a general sense of the layout of the phase diagram. This allowed us to find regions where we were highly confident about the correct classification, and we used these regions for training data. In the case of the microemulsion phase, this included checking that the correlation function had a local minimum.[d] After the first round of training and testing, we examined snapshots from different points in the phase diagram to visualize where errors occurred, and we added further training data at appropriate points to reduce these errors. For example, we initially did not include the crystal phase consisting of black and white rectangles, as this phase was not described in previous work. We identified this as a separate phase after it was labeled as fluid phase in earlier tests. The final training data set is shown in Figure S1a, overlaid on the final phase diagram. At each chosen set of training parameters (156 sets in total), 100 independent samples were acquired for training, for a total of 15600 samples in the training set.

Note that, despite the heuristic approach to generating the training data, it is not the case that we could generate an arbitrary different phase diagram simply by changing the training data. Rather, the phase diagram reflects real, qualitative differences in the behavior of the system. In our experience, training with a bad training set (e.g., containing different phases labeled as the same phase) leads to an obviously bad phase diagram, in which some regions contain different adjacent pixels classified as different phases with low confidence (quantified as described below).

Two types of training data were acquired for use in training two separate networks. In one data set (the snapshot approach, phase diagram shown in Figure S1c), simply 100 independent snapshots per parameter set were saved. In a second data set (the averaged approach, phase diagram shown in Figure S1b), 100 independent groups of 10 correlated snapshots each (as described in Monte Carlo Simulations, above) were acquired. The 10 snapshots were averaged to give one average image for the data set. Broadly speaking, this averaging has the effect of smoothing out random fluctuations, allowing the network training to focus on more constant aspects of each phase.

---

d    The appearance of this oscillation in the correlation function is one (admittedly somewhat arbitrary) definition of a microemulsion suggested by Gommper and Schick.[32]

## a  Classify snapshot phase with NN

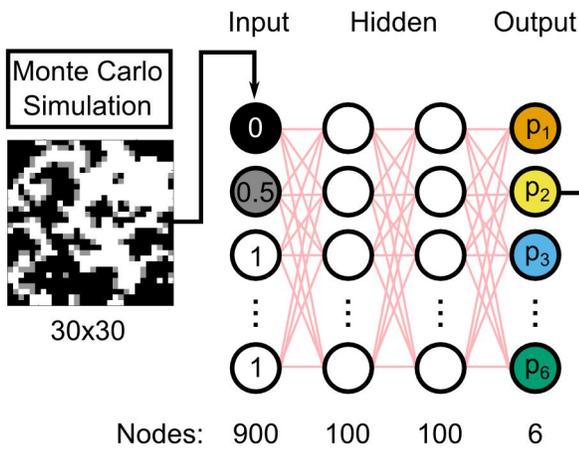

Nodes:  900    100    100    6

## b  Generate full phase diagram

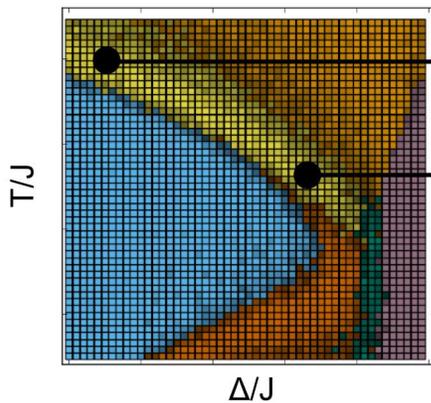

T/J vs Δ/J

## c  Calculate binding energies in different phase regions

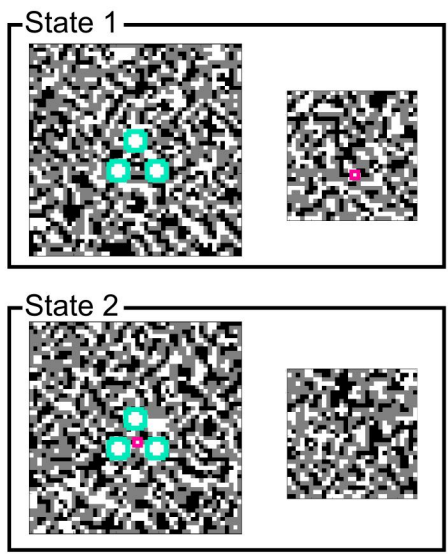

◻ = IgE Receptor
▪ = Lyn Kinase

**Figure 2**. Schematic of the methodology used in this study. (a) Schematic of the neural network (NN) used for phase prediction. The pixel values from a Monte Carlo simulation on a 30x30 lattice serve as inputs. (Black, white, and gray pixels are rescaled to non-negative values for these simulations as described in Methods). The network is trained using 2 hidden layers of 100 nodes each. The network contains 6 outputs, corresponding to its confidence that the input represents each of the 6 possible phases. Each pixel in the simulation box has dimensions of 1 nm x 1 nm. (b) At each point in parameter space (square pixels), the neural network was run on Monte Carlo simulation results to label the phase (c) Schematic of simulations used to calculate kinase binding energy by Bennett's method. The simulated system consists of two separate boxes, one representing the membrane near the receptor cluster (left), and another representing a section of the membrane at infinite distance (right). The teal and magenta proteins' $\sigma$ values are fixed white, while the rest of the lattices are Monte Carlo sampled. We use Bennett's method to calculate the free energy difference between State 1 (kinase at infinity) and State 2 (kinase inside cluster). The dimension of each pixel in the simulation boxes is 1 nm x 1 nm. The solid lines linking panel a to panel b and panel b to panel c show one example of how a phase is determined and used in Bennett's calculation.

The neural network code used is the implementation of [45], also available online at https://github.com/mnielsen/neural-networks-and-deep-learning . Each training sample was converted into an input vector of length 900 containing the values at each site of the 30x30 lattice, and a target output vector of length 6, consisting of 1 at the index of the correct phase, and 0 for all other values. The values of the input vector were rescaled such that black = 0, gray = 0.5, white = 1, in order to provide all non-negative inputs to the network (Figure 2a). The feed-forward neural network contained two hidden layers of size 100 each, made up of sigmoid neurons. We performed 25 epochs of training. In each epoch, the training data were randomly divided into mini batches of size 10. With each mini batch, stochastic gradient descent was performed by a backpropagation algorithm with a learning rate $\eta$ = 0.06. We use a cross-entropy cost function, with an L2 regularization parameter of $\lambda$ = 0.04 to avoid overfitting. To avoid stopping the stochastic training at a bad point, if the final classification accuracy was worse than 0.85, extra epochs were run, one at a time, until 0.85 was reached. For the snapshot approach, we instead used a threshold of 0.9. This method resulted in at most 5 (typically 0-2) extra epochs added. 10 instances of the neural network were trained independently on the same training data set. When working with the test data, we took the average output of the 10 instances.

### *Neural Network Phase Diagram Generation*

Test data were generated by the same Metropolis method as the training data. At each point in parameter space ($H$, $K$, $L$, $\Delta$, $T$; Eq. 1) where we sought to determine the phase, 5 snapshots or correlated averages were generated. These were fed as input into the neural networks, yielding output vectors with 6 elements in the range [0,1]. In these output vectors, a higher value at a particular index indicates that the point more likely belongs to the corresponding phase. Output vectors were averaged over the 5 samples and 10 network instances to arrive at a single final output vector (Figure 2a). The point was classified as the phase corresponding to the maximum value in the output vector (Figure 2b). The classification confidence was calculated as the maximum value in the output vector, divided by the sum of the output vector. When rendering the phase diagrams, the phase classification determined the color – red, green, blue, orange, pink, or yellow. The RGB value of the base color was multiplied by the classification confidence, such that a brighter[e] color represents more confident classification. For example, a point classified as lamellar (red, RGB = (0.8, 0.4, 0.0)) with confidence 0.8 would be rendered as RGB = (0.64, 0.32, 0.0).

The averaged approach was more effective than the snapshot approach. With the snapshot approach, we could only distinguish 4 phases: Fluid, Black/White, Gray, and a single region covering Lamellar, Microemulsion, and Critical (Figure S1c). With the averaged approach, we could distinguish six phases (Figure S1b), but we had low confidence in the distinction between the Fluid and Gray phases (Figure S1d). To combine these, on testing

---

e Note that we use the term brightness here in the sense of the HSB (Hue, Saturation, Brightness) representation of colors. HSB and HSV (Hue, Saturation, Value) are equivalent representations, so scaling the brightness is synonymous with scaling the value.

data, we used the Gray output from the snapshot approach, and the other 5 outputs from the averaged approach. This gave the final phase diagram that we believe most completely describes our understanding of it after our work with both these approaches.

*Binding Energy Computation*

We consider the binding energy to be the difference in free energy between a single white pixel (spin +1) with a set cluster of three other white pixels, compared to that single white pixel being at an infinite distance from that set cluster (Figure 2c). We call the set cluster "receptors" and the designated single pixel, "kinase." To compute this binding energy by Bennett's method,[46] simulations were performed on the four separate lattices shown in Figure 2c: State 1 consists of a 50x50 lattice containing the set cluster of receptors, and a separate 30x30 lattice containing the kinase. State 2 consists of a 50x50 lattice containing the kinase within the cluster of receptors, and a 30x30 lattice empty of the kinase. Note, for the 30x30 boxes (Figure 2c, right), the smaller lattice size was permissible because these boxes only ever contain one designated white pixel, which affects the lattice on a shorter length scale than the full receptor cluster. Samples were generated by the Metropolis algorithm in the same way as the neural network training data, but the predefined receptor and kinase proteins were required to remain white. Any proposed move that attempted to flip one of these spins was automatically rejected.

The free energy $\Delta F$, corresponding to the binding energy, is computed according to the following formula.

$$e^{-(\Delta F - C)/(k_B T)} = \frac{\langle f((\Delta U_{1 \to 2} - C)/(k_B T)) \rangle_1}{\langle f((\Delta U_{2 \to 1} + C)/(k_B T)) \rangle_2} \quad (2)$$

Here, $C$ can be any constant, with the fastest convergence achieved when $C \approx \Delta F$. We choose $C = -0.5\ k_B T$, and choose $f$ as the Fermi-Dirac function, $f(x) = 1/(1 + e^x)$ as suggested in [46]. The numerator is calculated as an ensemble average from simulations of state 1 (Figure 2c, top). $\Delta U_{1 \to 2}$ for each sample is the energy change associated with exchanging the kinase and a pixel at the center of the cluster (corresponding to the kinase position in state 2). Likewise, the denominator is calculated from simulations of state 2, and $\Delta U_{2 \to 1}$ is the energy change associated with exchanging the kinase located within the cluster and the pixel corresponding to its position in state 1.

Note that the two separate boxes that make up each state in Figure 2c can be generated independently, and we use this to our advantage. We initially generated the same number of samples of the 50x50 box and the 30x30 box. Then each 50x50 box was paired with 10 different 30x30 boxes, increasing the number of samples of the state by a factor of 10. These samples are not independent, but they still follow the correct Monte Carlo statistics.

For calculating binding energy at each parameter set to be tested, simulations were performed for 5000 sweeps, a sample was saved every sweep, and the lattice was reshuffled every 10 sweeps. After data expansion, this gave 50000 non-independent samples of each state, to be used in the Bennett calculation.

## Results
### *Neural Network Phase Identification*

We trained neural networks to classify the output of a Monte Carlo simulation of the Gompper and Schick lattice model,[35] according to the phase that the simulation represents. A schematic of the network and an example of a resulting phase diagram are shown in Figure 2a, b. The $\sigma$ values from a 30 by 30-pixel Monte Carlo snapshot (generated by the standard Metropolis method)[47] were used as 900 inputs to the network. Training data consisted of 15600 such snapshots, which represented typical examples of each phase of interest (Figure S1). The network was trained with 2 hidden layers of 100 nodes each, and an output layer of 6 nodes, corresponding to the six phases of interest in the phase diagram. Alternatively, instead of single Monte Carlo snapshots, we used input consisting of the average of 10 correlated snapshots from consecutive simulation steps. This method tended to be more accurate in most cases, and our final reported phase diagrams make use of some output from both types of networks. Our procedures, including training of the neural networks, are further described in Methods.

We initially evaluated the lattice model with $H/J = 0$, $K/J = 2$, $L/J = 3$, ranging over $T/J$ and $\Delta/J$ values of order 1. In the original description of the model,[35] this cross-section was found to contain examples of all phases present in the model.

Our neural network was able to confidently label six distinct regions of the phase diagram (Figure 3), corresponding to the eight phases described in the Introduction: fluid, lamellar, gray, crystal, black / white, and microemulsion / critical. The network was not able to determine a distinct boundary between microemulsion and critical fluctuations, so the single microemulsion / critical label was applied to both. At larger values of $\Delta$, the region is a microemulsion, while, at smaller values of $\Delta$, the system shows fluctuating domains consistent with close proximity to an Ising critical point. Instead of a clear boundary between the critical fluctuation and microemulsion behaviors, the regions blend into one another smoothly. Because two models often used to explain lipid rafts – microemulsion and critical phenomena[1] – are included within this phase, it is highly relevant for membrane-related questions.

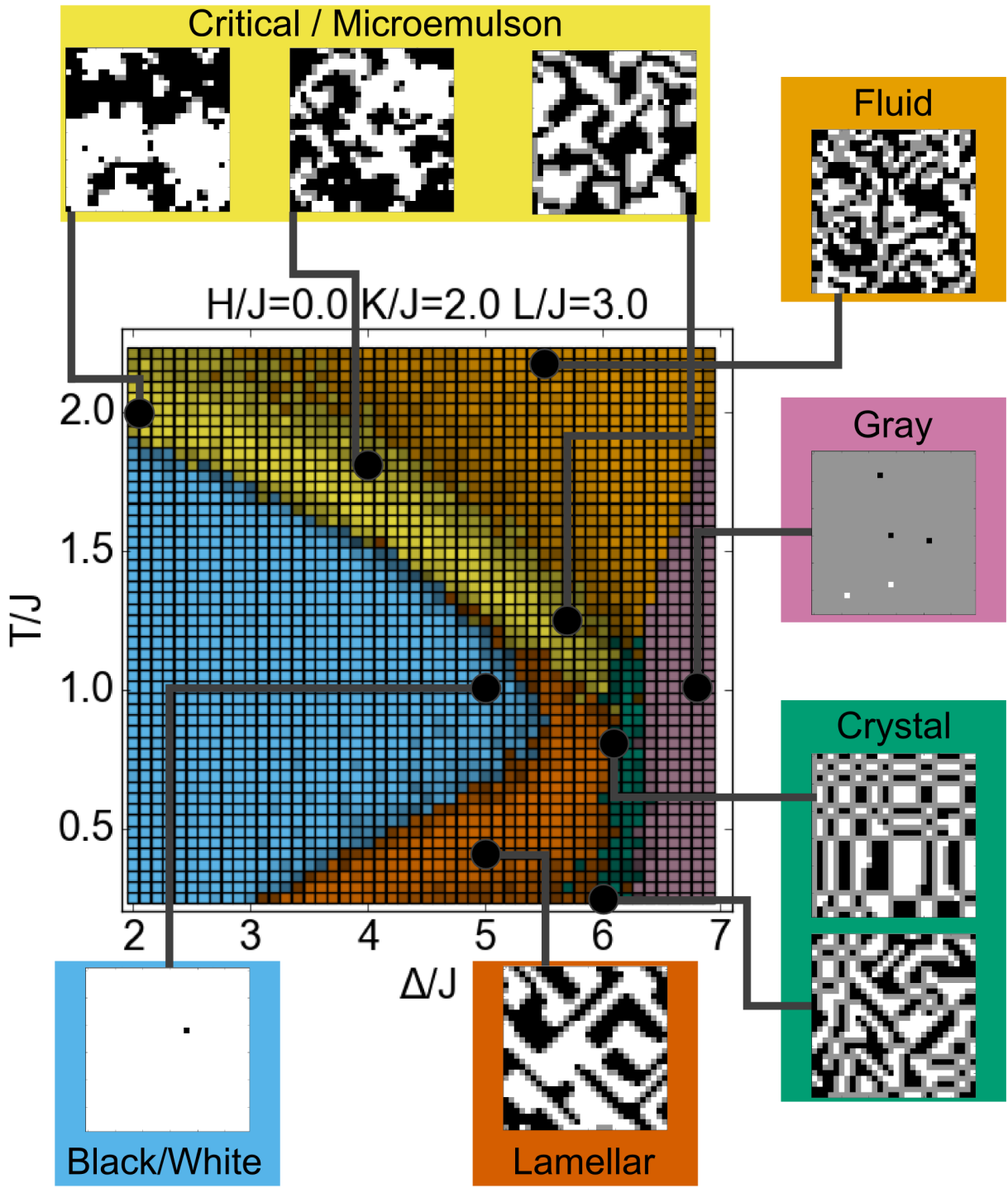

**Figure 3**. Phase diagram of the lattice model. The color of each pixel with specified (*Δ/J*, *T/J*) coordinates indicates the phase at that point, as determined by the neural network. Pixels with a higher brightness indicate a higher level of confidence in the classification. Snapshots show typical examples of each of the phases, corresponding to the black points on the phase diagram.

The network applied the same black / white label to both the black phase and white phase. Because the training data contained examples of two-phase coexistence, including snapshots of both black phase and white phase with the same classification, the neural network was trained to apply the same label to both. With $H$=0, the black / white classification represents two-phase coexistence between the black and white phase, while with $H$>0, the white phase does not exist, and the label represents only the black phase (conversely for $H$<0). Finally, we note that the network applied the crystal label to the limit of the lamellar phase in which the components alternate with period of one lattice unit.

### *Exploring the Phase Diagram*

We used our neural network to compute other cross-sections of the phase diagram and thereby gain a more complete perspective on the entire parameter space. Remarkably, it was not necessary to retrain the network to work with these other cross-sections. We found that the original network trained at $H/J$=0, $K/J$=2, $L/J$=3 accurately identifies the phases elsewhere in the phase diagram, for all $H$, $K$, and $L$ values considered in this study.

Varying the surfactant strength $L$ changes the topology of the phase diagram (Figure 4). At zero or low $L$ ($L/J$ = 1.5), the lamellar phase does not exist, and the black/white phase directly borders the gray phase. At zero $L$, a tricritical point exists at the intersection of the fluid, black/white and gray phases. At higher $L$ ($L/J$ = 3), we reach the case shown in Figure 3, in which the lamellar and crystal phases exist between the black/white and gray phases. At even higher $L$ ($L/J$ = 6), the system becomes a crystal for nearly all values of $\Delta$ and $T$ tested, maximizing the number of surfactant interactions.

With $K$=0 and $L$=0, the model reduces to the more widely studied Blume-Capel model (Figure 5), in which gray pixels are neutral in their interactions with white and black pixels. In our diagram, the region between black/white coexistence and the fluid phase can be identified as a critical transition by virtue of the yellow critical region appearing between the blue and orange regions. Note that microemulsions are not possible with $L$=0, and therefore the entire yellow region in this cross-section represents Ising critical behavior. The critical line occurs at the boundary between the blue and yellow regions in Figure 5a. With $H/J$ = 0.1, the critical "phase" disappears, correctly showing that at $L$=0, $H$>0, there is no longer a critical

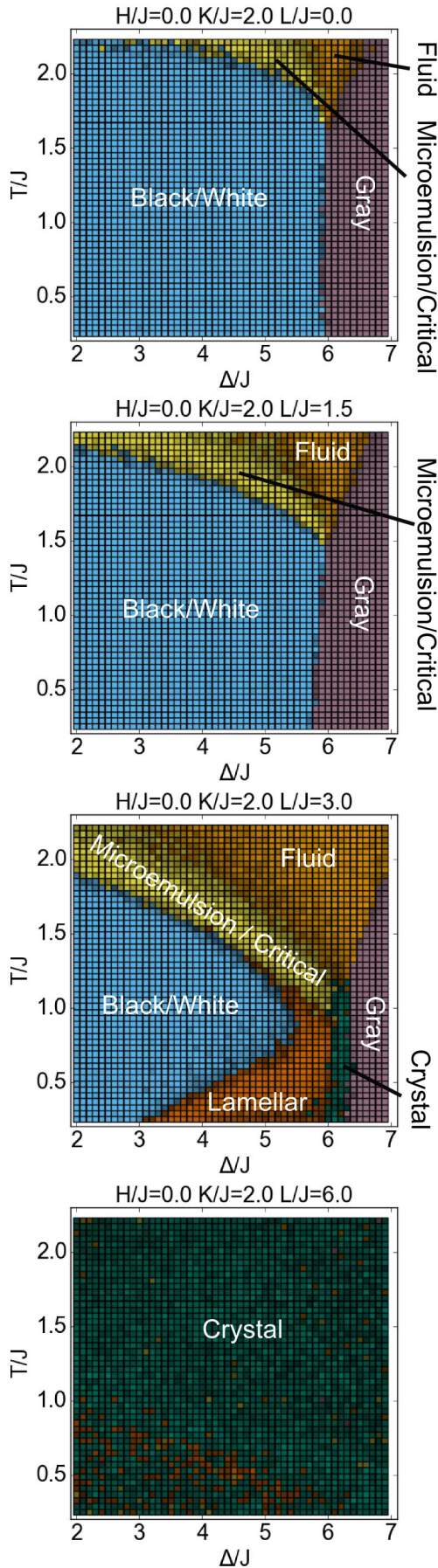

**Figure 4**. Cross-sections of the phase diagram at varying values of the surfactant strength. Surfactant strength $L/J$ is varied 0 to 6, with constant $H/J = 0$, $K/J = 2$. Colors have the same meaning as in Figure 3.

phase transition (Figure 5b).

When $L$ is increased with $K=0$ (Figure S2), the phase diagram has topology similar to as the case with $K/J = 2$, although the phase boundaries occur at lower $\Delta$. Finally, we considered some additional cross-sections at positive $H$ (Figure S3). We note that with $H/J=0.1$, $K/J=2$, $L/J=3$, some yellow region remains at high $\Delta$. Presumably this indicates microemulsion behavior, because a critical line is not expected to exist at nonzero $H$. At higher $H$ ($H/J=0.5$), the black/white classification (here representing only the black phase) grows to encompass most of the parameter space examined in this range of $\Delta$ and $T$.

### Quantifying Protein Recruitment in Terms of Preferential Partitioning

Having calculated the phase diagram for the lattice model, we turned to our questions related to biological function. In particular, we compare the effectiveness of lipid raft-mediated protein reorganization at various points on the phase diagram. As a specific test, we consider the case of three receptors (such as IgE-FcεRI) cross-linked to form a cluster; these are activated to initiate transmembrane signaling only after recruiting a membrane-anchored kinase (such as Lyn; see Figure 1). We assume both the receptors and the kinase prefer Lo-rich domains (i.e., lipid rafts), and correspondingly, we represent them with white pixels, which we place at selected, fixed positions in the lattice. Lyn, is represented by one white pixel, whereas each of the three receptors is represented by 12 white pixels, corresponding to their relatively larger size (Figure 2c). We calculate the binding energy as the free energy associated with moving the kinase into the middle of the three-receptor cluster. A larger magnitude negative value indicates a stronger contribution of lipid rafts to protein colocalization at a particular point in the phase diagram.

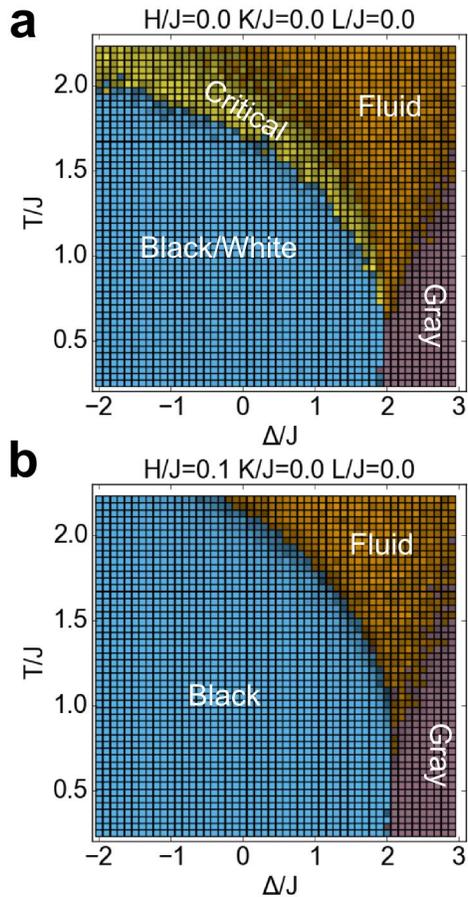

**Figure 5**. Cross-sections of the phase diagram for the Blume-Capel model (*K*=0, *L*=0), with *H*/*J* = 0 or 0.1. Colors have the same meaning as in Figure 3.

Similar to what we and others have done previously,[44] we use Bennett's method[46] (Eq. 2) to calculate the free energy change. We do so here in a more computationally efficient method than in previous studies. In previous work, we calculated the energy change stepwise, moving the kinase out of the cluster, one lattice unit at a time, and generating a profile of energy versus position in the process.[44] Here, we instead calculate the entire energy in one step. Our simulated system (Figure 2c) consists of two separate boxes, one containing the receptor cluster (left), and the other representing membrane at infinite distance from the cluster (right). By Bennett's method, we compute the free energy to move the kinase from the box at infinite distance (state 1) to the center of the cluster (state 2).

We used the phase diagram to assist in choosing points for Bennett simulations – we ran a simulation at each point marked with a diamond in Figure 6. We focused our simulations primarily on the microemulsion/critical region of the phase diagram, and, for comparison, we performed simulations at a smaller number of points elsewhere in the phase diagram. We additionally performed simulations in which a single white pixel was set (instead of the receptor cluster) and calculated binding energies for a second white pixel to come into proximity. We found these binding energies to be qualitatively similar but weaker compared to the case with the cluster (Figure S4).

Our results for a kinase associating with a receptor cluster (Figure 2c) are shown by the colors of the diamonds in Figure 6. In the Blume-Capel ($H = K = L = 0$) phase diagram (Figure 6a), no microemulsion exists and we find roughly the same binding energy of ~ -0.6 $k_B T$ at all points along the Ising critical line at the boundary between the blue and yellow regions. This corresponds to a modest increase in kinase concentration, by a factor of $e^{0.6} \approx 1.8$. Along this critical line, the binding energy does not show a dramatic difference above, versus below, the transition temperature (columns of diamonds along blue-yellow boundary). Strikingly, as the tricritical point is approached (the box in Fig. 6a), we find a dramatic, nearly threefold increase in the magnitude of the binding energy. The minimum free energy of -1.7 $k_B T$, is achieved at the tricritical $\Delta$ (1.9655 $J$) and 1.04 times the tricritical temperature (0.634 $J/k_B$). The corresponding increase in kinase concentration by a factor of $e^{1.7} \approx 5.5$ is much more significant than the 1.8 factor at an ordinary critical point. We suspect that the distance of the optimum above the tricritical temperature, 1.04x, is a finite size effect, as this value increases if the simulation box is made smaller. The true optimum might occur at exactly the tricritical temperature (0.610 $J/k_B$).

To validate our new application of Bennett's method (Eq. 2, Figure 2c), we also calculated the energy profile at the tricritical point stepwise by Jarsynski's method,[48] identical to the method used in [44] (Figure S5). Due to the larger simulation box used in this method, finite size effects are less of a concern. We found a binding energy of ~ -1.5 $k_B T$ with Jarzynski's method, comparable to our result at 1.02x the tricritical temperature with Bennett's method (Figure 6b). However, at the tricritical temperature, our application of Bennett's method gives a binding energy of only -1.0 $k_B T$, presumably due to finite size effects at this temperature.

We compare these results to the first-order phase transition that occurs at $H > 0$ (Figure 6c), which yields a higher concentration of black (Ld-preferring) pixels than white (Lo-preferring) pixels in the lattice. We found a similar binding energy of ~ -0.6 $k_B T$ above the transition temperature in the fluid phase. However, we see a substantially stronger binding energy as low as ~ -1.4 $k_B T$ upon entering the phase-separated state. In the context of membranes, this would correspond to a situation in which most lipids on the membrane favor the Ld phase, but our receptor/kinase proteins of interest favor Lo.

Finally, in Figure 6d, we consider the binding energy around the microemulsion/critical region using the parameters of Figure 3 ($H/J=0.0$, $K/J=2.0$, $L/J=3.0$). To aid in the distinction between microemulsion and critical "phase" in this cross-section, at selected points (marked

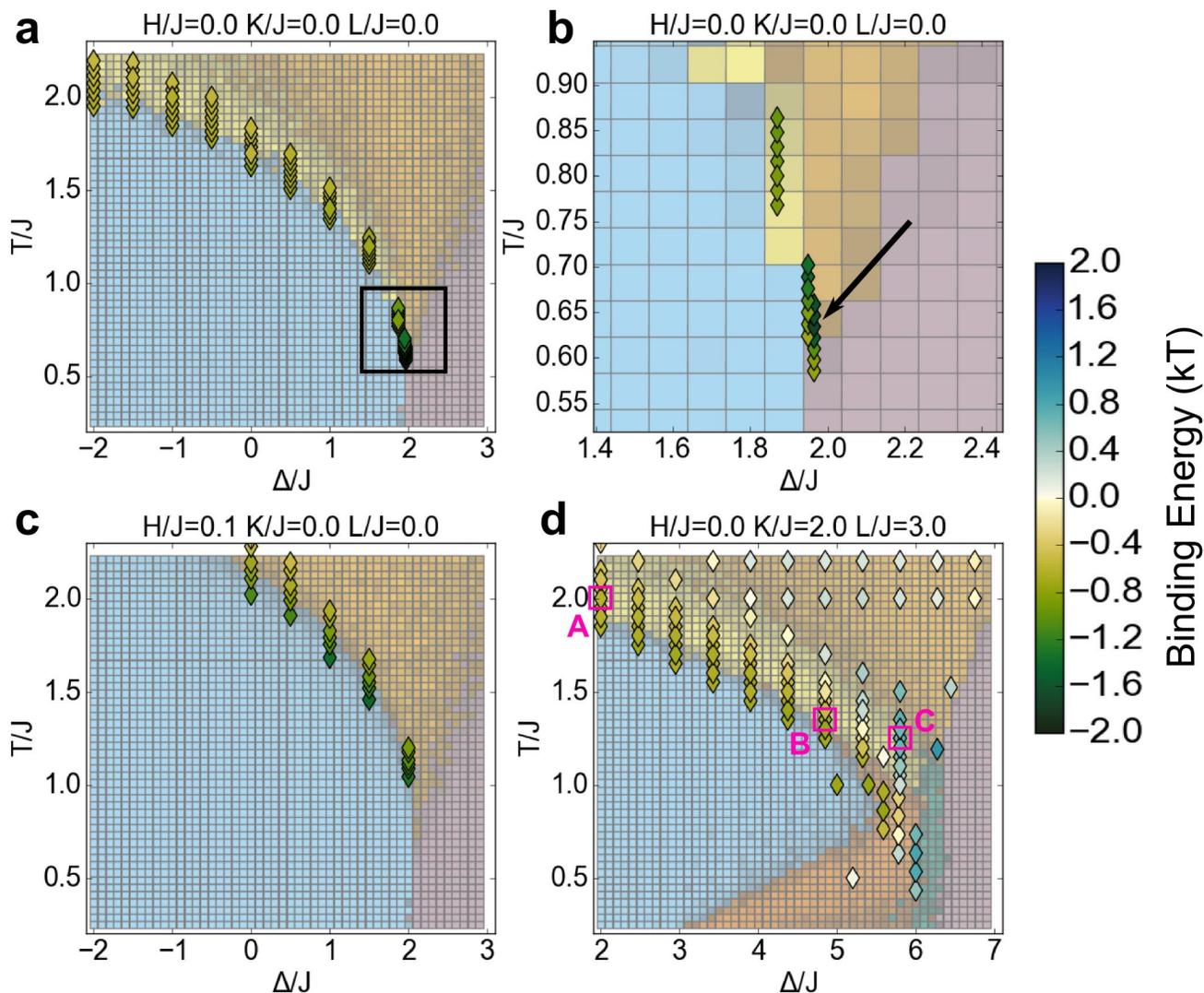

**Figure 6**. Kinase binding energy at selected points in the phase diagram. Each colored diamond indicates the free energy change associated with moving a kinase into a cluster of receptors (as in Figure 2c) at that point ($\Delta/J$, $T/J$) in the phase diagram. The phase diagram colors are rendered paler than in other figures to make the diamonds more clearly visible. (a) Binding energy in the Blume-Capel model ($K=0$, $L=0$) for the cross-section $H=0$. (b) Inset of (a) in the region around the tricritical point (black box in (a) ). The indicated point (black arrow) with the minimum free energy of -1.7 $k_BT$ occurs at the tricritical $\Delta$ and 1.04 times the tricritical temperature. (c) Binding energy in the Blume-Capel model for the cross-section in which the external field $H/J = 0.1$ favors black pixels, opposite to the kinase and receptor preference for white pixels. (d) Binding energy in a cross-section that includes microemulsions and lamellar phases ($K/J = 2$, $L/J = 3$). Within the microemulsion/critical phase, marked points (magenta boxes) were analyzed with correlation functions and visual inspection of simulation snapshots (Figure S6). Point A is part of the critical "phase," point B is a microemulsion with length scale ~ 10, and point C is a microemulsion with length scale ~ 4. At certain points in this cross-section (blue color scale), including point C, the positive binding energy indicates that it is energetically unfavorable to bring the kinase into the cluster.

A, B, and C in Figure 6d), we performed correlation function analysis (Figure S6). We confirmed that point A is in the Ising critical region, and B and C are in the microemulsion region. In the Ising critical region (i.e. the yellow region at low values of $\Delta$, including point A), we again find a binding energy of ~ -0.6 $k_BT$, the same as the case with no surfactant strength L (Figure 6a). As we move to higher $\Delta$, corresponding to a microemulsion region, we find a striking change. At a subset of the points in the microemulsion region (including point C), the binding energy at becomes much weaker, even turning positive (unfavorable). Intuitively, this happens when the characteristic length scale of the microemulsion is smaller than the size of the set receptor cluster. Considering microemulsions with a longer length scale (near blue-yellow boundary at $\Delta/J$ between 4 and 5, including point B), we find a binding energy of -0.6 $k_BT$, comparable to that at an Ising critical point at lower values of $\Delta$. Thus, the results indicate that the binding energy associated with microemulsion behavior depends on how the characteristic length scale of the microemulsion compares to the spacing of the clustered receptors. It is also possible that a microemulsion exists at a length scale larger than our 30x30 nm snapshot used to generate the phase diagram. This would likely appear as phase-separated in our diagram, and indeed would look equivalent to phase-separated from the perspective of a cluster of size less than 30 nm. Based on our simulated results at points in the phase-separated region, this case would also likely yield a value of around -0.6 $k_BT$ (Figure 6d).

The Figure 6d cross-section contains no point comparable in binding energy to the tricritical point in Figure 6a. The minimum binding energy achieved in this cross-section (except perhaps at biologically irrelevant points at very low temperature) is ~ -0.6 $k_BT$, which occurs along the entire boundary between the black/white and microemulsion/critical regions. This remains true if we more densely sample the entire length of the phase boundaries (data not shown). Among all of the phase states tested, the tricritical point at $H = K = L = 0$ (Figure 6a) leads to the strongest possible binding energy for kinase and clustered receptors.

**Discussion**

*Comparison to Published Results*

We have generated the phase diagram for the Gompper and Schick lattice model using relatively new neural network methodologies. It is important to consider how this method compares to other more established methods for phase diagram determination. We examine certain special cases of the model that allow for direct comparison of our phase diagram to published phase diagrams obtained by other methods.

By taking $\Delta$ to -∞ (no gray pixels), and $H=0$, we have the Ising model, with the well-known critical transition temperature of 2 / log(1 + sqrt(2)) ≈ 2.269 $J/k_B$. Applying our existing network to this case, we see the phase transition at close to the correct temperature (Figure S7a). The network's confidence level for the Ising model is worse than optimal because this network was trained to perform a more complicated classification on 6 phases, instead of the 2 phases (fluid and black/white coexistence) relevant to the Ising model. A different approach is to train a network, solely on Ising model examples, to classify between only the fluid phase and black/white coexistence. With this model, we distinguish the phases with high confidence, and we nearly perfectly identify the transition temperature (Figure S7b). This level of accuracy is comparable to previous neural network work on the Ising critical transition.[41]

The result for the Blume-Capel model ($K=0$, $L=0$; Figure 5) with $H=0$ is comparable to results with this model from other methods. We find good quantitative agreement on the location of phase boundaries with Beale's phase diagram from finite size scaling[49] (Figure S8). We also show the mean field theory solution[37] for comparison. The tricritical point has an upper critical dimension of three, meaning that mean field theory is expected to be inaccurate near the tricritical point in this two-dimensional model.[50] However, our calculated result is much closer to the more accurate finite size scaling solution.

Our diagrams can also be compared to those obtained in Gompper and Schick's original description of the model[35] (Figure S9). Note that, to make this comparison it was necessary to add the parameter $K2$, the equivalent of $K$ between second nearest neighbors in a straight line. This had no effect on the overall shape of the phase diagram, but shifted the phase boundaries slightly. We find very good agreement on the location of the critical line in all cross-sections with Gompper and Schick's transfer matrix approach. The original phase diagram included a Lifshitz line, which the authors defined as the separation between Ising and microemulsion regions. This helps us better interpret the combined microemulsion/critical region in our phase diagram, which is in fact a microemulsion to the right of the Lifshitz line. In other aspects of the phase diagram, the neural network approach provided new information, and it revealed shortcomings of the original phase diagram. We note our new placement of the lamellar phase (red) is qualitatively different from the Gompper and Schick diagram, including a lobe that sits below the phase-separated state on the temperature axis. We give a new boundary between the gray phase (pink) and the fluid phase (orange). Our identification of the rectangular crystal phase (green) is entirely new, not addressed in the original study (the diagonal crystal that our network labeled as part of this phase arguably belongs in the lamellar phase, but the rectangular features are clearly a distinct phase).

Some of these novel features are relevant to the biological system of interest, while others are not (such as the rectangular crystal phase, which likely exists only due to the use of a square lattice), but all point to the strengths of global computational approaches in phase diagram prediction, which allow direct comparisons. Theoretical techniques like finite size scaling frequently focus on specific interesting areas of the phase diagram, such as the critical line. In our neural network approach, we instead indiscriminately analyzed entire slices of the phase diagram, extracting features in both critical and non-critical regions. This is especially valuable for a problem such as biological lipid-based membranes, for which different groups have proposed that the most relevant states are either near a critical point[21,29] or away from a critical point.[7,26,28]

Finally, to further validate the application of this model to the study of lipid membranes, we compare our neural-network-derived phase diagrams with the numerical and mean-field phase diagrams produced in previous studies on the formation of lipid rafts (nanoscale domains concentrated with Lo-preferring components, as defined in the Introduction). We consider first microemulsion-based models, which propose that either surfactant-like lipid species[28] or membrane curvature[7,26,27] stabilize the interface between different phase domains. Importantly, the generality of our neural network approach means that we could in principle explicitly reproduce the results of the different membrane models described above. It should even be possible to train a neural network with multiple models simultaneously, a potential avenue for future work. Here, however, we are interested in comparing the results of our single-Hamiltonian neural network approach with results in the membrane modeling literature.

How much agreement should we expect between the neural network trained on our Hamiltonian (Eq. 1) and models with different Hamiltonian forms and explicitly different energetic terms (e.g. composition-curvature interactions)? Due to the presence of gray pixels as surfactants, our model most closely resembles models that make use of hybrid lipids,[28,51] so we can ask how our model compares to curvature-based models, which are seemingly the most different. As discussed above (Results), our Hamiltonian captures much of the physics of other membrane models, including 2D Ising critical and tricritical behavior. In these critical regimes, our Hamiltonian is equivalent to all others due to the universality of critical behavior.[50]

Outside these critical regimes, in the biologically-relevant microemulsion phase, we also expect qualitative agreement between our model and curvature-based models. Intuitively, in a microemulsion regime, the gray pixels in our model will act analogously to regions of curvature mismatch: in a system with droplets of one phase suspended in a backdrop of another phase, the boundaries of the droplets will be regions of concentrated surfactant-like interaction. In our model, this looks like a domain of either black or white pixels encircled by a strip of gray pixels; in the curvature model, the picture is the same, except that the gray pixels are replaced by a region of curvature change (this can be pictured as the droplet "popping out" of the membrane). Importantly, in this regime we have a defined length scale in both models: in ours, it arises from the concentration of gray pixels, while in the curvature model it arises from the mechanical properties of the membrane.

Ultimately, however, the comparison of predicted phase behavior serves as the best indicator of model similarity, and we find good agreement between the phase diagrams in the literature (both curvature- and hybrid-lipid-based) and those generated by our neural network approach (Figure 7). Our phase diagrams reproduce all the features found in these other model frameworks, including Ising critical transitions, lamellar phases, two-phase coexistence and tricritical phenomena. Moreover, the general topology of the phase diagrams is consistent regardless of model choice—for instance, all models considered here predict a lamellar phase separated from a microemulsion phase by an Ising critical line, with the microemulsion phase, in turn, separated from an ordinary fluid phase by a boundary that is not a true thermodynamic phase transition. This consistency with previously calculated phase diagrams[7,51] speaks to the generality of our approach, which allows us to describe and compare a wide variety of membrane phenomenologies using a single model framework.

*Application to Lipid Rafts*

We set out with this model to analyze competing hypotheses on the physical basis for formation of lipid rafts: does stabilization of nanoscale Lo-like domains arise from proximity to a critical phase transition, or from nanodomains of a characteristic size, as in a microemulsion? We found that in some ways, the two hypotheses are much alike. As described in the Introduction, considerable evidence supports the view that lipid rafts serve to recruit proteins to the correct place on the cell membrane, such as our example of Lyn kinase recruitment into a set IgE-FcεRI cluster, where both components are Lo-preferring. Our phase diagram shows that critical and microemulsion phase states can be equally beneficial thermodynamically for this membrane purpose. As we showed, both can give about the same optimal binding energy of -0.6 $k_BT$. We also showed it is possible to sit in a region between microemulsion and critical point with a classification that is subjective. Gompper and Schick chose the Lifshitz line as an arbitrary distinction for what qualifies as a microemulsion, while our neural network was unable to draw a sharp line between the two behaviors. Our energy calculations make a clear prediction for a difference between clearly critical and clearly microemulsion states (at lower and higher values of $\Delta$, respectively). Microemulsions carry the requirement of a particular characteristic size, and can only effectively stabilize lipid domains smaller than that size. If the set cluster of Lo components is larger than the microemulsion length scale, then there is actually *exclusion* of other Lo components from the cluster (Figure 6d). In contrast, if the membrane sits near an Ising critical point, the consequent lipid rafts are stabilized at all length scales, never excluding other Lo components. If the membrane indeed exists as a microemulsion, then in principle it should be possible to experimentally exceed the correct length scale, and cause a reversal of the lipid mediated signaling. To our knowledge, this exact experiment has not been carried out, and may remain challenging to implement. However, in mast cells, a structurally defined ligand with spacing 13 nm has been studied,[52] and the resulting large receptor spacing lowers, but does not eliminate, the signaling response. This suggests that, if the mast cell signaling response relies on a microemulsion-mediated kinase recruitment, that microemulsion length scale must be larger than 13 nm.

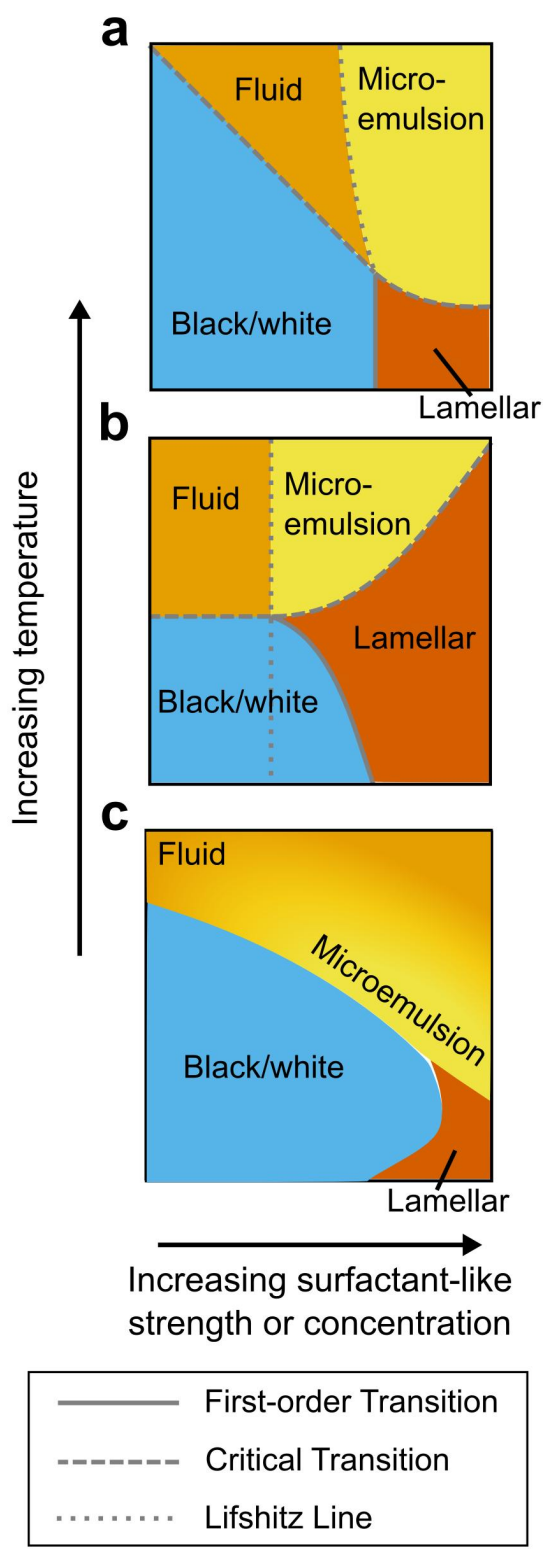

**Figure 7**. Schematic comparison of our phase diagram with those from the microemulsion literature. (a) Mean-field phase diagram from a model with hybrid lipid species acting as a surfactant. Adapted from [51] (b) Mean-field phase diagram from a model with curvature coupled to membrane composition to produce a surfactant-like interaction. Adapted from [7]. (c) Phase diagram generated by our neural network approach. X coordinate gives strength of surfactant-like interaction in (b), or concentration of surfactant species in (a) and (c). Y coordinate represents temperature. Note that this is a schematic representation, so the actual axes from the source papers differ in scale and representation. For the sake of comparison to the other models, we use yellow here to represent only microemulsions, not Ising critical behavior. The yellow-orange gradient in (c) is used to schematize the ambiguity between microemulsion and ordinary fluid phase, and represents our best interpretation of the location of the microemulsion state, taking into account the neural network output (Figure 3), snapshots within the phase diagram (Figure 3, yellow-bordered panels), and the location of the Lifshitz line from [35] shown in Figure S9.

One argument sometimes used in favor of microemulsions is that they are easier to achieve, requiring less cell-directed tuning of the membrane. However, our phase diagram points to an additional complication: the cell not only has to tune the membrane composition to a microemulsion, but also must tune the length scale to the characteristic size necessary for the correct biological function, which may be highly variable, depending on the signaling pathway and components involved.

What about the actin cytoskeleton? It is widely thought that cortical actin couples to the membrane, forming "corrals" that add further complexity to the heterogeneity of the membrane. However, in many ways this does not affect our conclusions, as typical size estimates for actin corrals[53] are above our simulation size of 50x50 nm (Figure 2c). A small cluster of Lo-preferring components set within one corral sees a particular membrane composition, regardless of the corrals boundaries at longer length scales. However, actin involvement motivates two other considerations. First, we should not ignore the phase-separated region of the phase diagram (blue, Figures 3-5, S2-S3), as the membrane may have a phase-separating composition, driven below the diffraction limit only by actin-mediated partitioning.[12,54] We see that a phase-separating membrane would yield a kinase binding energy similar to the minimum in the microemulsion/critical phase. (Figure 6a,d). Second, we note that, due to cortical actin, the membrane composition encountered by receptors might not be the global composition of the membrane. This actin meshwork has been proposed to preferentially sequester either Lo or Ld lipids,[12,34] which would deplete these from a cluster set in the middle of a corral.

The most striking new discovery from our phase diagram and energy calculations is the power of a membrane at a tricritical point. Our computations show that near the tricritical point, the potential binding energy due to lipid rafts increases by a factor of 3 compared to any of the other proposed models: critical point, microemulsion, or phase-separated two-phase coexistence. This increase in energetic favorability could confer significant advantage for lateral recruitment in the membrane.

Moreover, the interactions near a tricritical point are long-range in nature, which could also have important implications for signaling in the natural cell environment. Previous conformal field theory results have shown that, near an Ising critical point, proteins can interact via long-range critical Casimir forces.[44] Because our method of calculating binding energies agrees with the results from [44], our system should also exhibit such long-range interactions near the tricritical point, but with a different power law governing their spatial decay. The large effect at the tricritical point likely comes from the different critical exponents of this universality class. In particular, the potential that gives rise to critical Casimir forces scales with the correlation function $g(r)$, which itself scales as $g(r) \sim r^{-d+2-\eta}$ near a critical point, where $d$ is the dimension of our system and $\eta$ is a universal critical exponent.[50] Plugging in the relevant critical exponents, we see that $g(r)$ scales as $r^{-0.15}$ at the Ising critical point, but as $r^{-0.25}$ at the tricritical point.[50,55–58] Thus, at the tricritical point, the critical exponent ($\eta=0.15$) allows attraction between Lo-preferring components to remain stronger at a longer distance, especially compared to direct chemical bonds or electrostatic interactions (which are expected to be screened over ~1 nm in the cell), and also longer range than the $r^2$ attractive forces mediated by membrane curvature.[44]

To our knowledge, a tricritical point has not previously been considered as a serious proposition for the physical basis of lipid rafts, and perhaps for a good reason: achieving proximity to a tricritical point requires tuning of three relevant parameters, whereas proximity to an Ising critical point requires only two. In the three-dimensional phase space of the Blume-Capel model, only a single point is a tricritical point. However, we note that in a many-component cell membrane with many more than 3 degrees of freedom, there would be more possibilities for tuning to a tricritical composition. The detailed nature of "lipid rafts" is probably quite variable even within a single functional cell membrane, and localized tuning may be possible for a particular signaling purpose. Furthermore, we argue that if effective lipid rafts provide a strong enough evolutionary advantage for the cell to respond appropriately to environmental stimuli, it might be to the cell's advantage to maintain a tricritical composition (at least locally), and gain the massive improvement in lipid raft energetics that results. Conversely, the optimal lipid raft strength for signaling to be appropriately regulated in the cell might be weaker than what is generated by the tricritical point, in which case we would expect the membrane to exist in one of the other phase states explored in this study.

It is also reasonable to ask whether lipid rafts could facilitate interactions between Lo-preferring components that lead to the formation of the cluster itself. This was not the case we considered for mast cells, in which clustering was due to physical cross-linking of the IgE-FcεRI by antigen. In T cell receptor signaling, for example, clusters form in the absence of cross-linking by a mechanism that remains unclear.[59] The Ising critical point or microemulsion binding energy of -0.6 $k_BT$ would not be sufficient to cause clustering of individual receptors; this requires considerably stronger interactions. We previously performed calculations and simulations based on the formulas for Casimir forces given in [44] and concluded that these forces, at an Ising critical point, are not large enough to mediate receptor clustering (Milka Doktorova and Eshan Mitra, unpublished observations). However, we now note that the stronger binding energies found near the tricritical point may be sufficient to mediate receptor co-clustering, independently of external agent.

We further note that the concept of a membrane at a tricritical point is not inconsistent with observations of GPMVs showing ordinary Ising critical exponents.[21] We argue that a membrane might exhibit tricritical behavior at short length scales and Ising critical behavior at the longer length scale accessible with current experimental techniques. This hypothesis can be formalized using renormalization group (RG) theory, a tool for describing how the observed behavior of a system changes due to coarse-graining. Here, coarse-graining corresponds to the loss in resolution when a membrane is observed with a diffraction-limited microscope. Certain points in parameter space are RG fixed points, which are unaffected by coarse-graining (i.e. look the same at different length scales). Other points, under RG coarse-graining "flow" towards or away from the fixed points (Figure S10). The 2D tricritical fixed point and Ising critical fixed point are two examples of such RG fixed points, with systems tending to flow from tricritical to Ising behavior under coarse-graining. As seen in Figure S10, physical systems that flow near to the tricritical point will show tricritical behavior on length scales relevant for protein organization, but could then flow away to Ising behavior on the longer length scales observed in GPMV studies.[f]

While our work with this lattice model has been useful in addressing many hypotheses on lipid organization (and proposing a new, tricritical possibility), it has some limitations. In

---

f   Indeed, the phase diagram of a physical system near a critical point echoes the flow diagram near the corresponding renormalization-group fixed point (the irrelevant contracting directions only making analytic changes in the phase boundaries), leading to a common conflation of the two (adding 'flow' arrows to the boundaries in experimental phase diagrams).

particular, this is a thermodynamic model, operating under the assumption of a steady state. Kinetic hypotheses about lipid organization, such as active actin remodeling,[60] would require a different theoretical framework in order to compare to the cases that we have explored. However, our neural network-based methods should allow similar morphological classification. Moreover, while it is possible for active processes to be described by Ising critical behavior,[61] studies on GPMVs isolated from cells[21] show that these membranes remain close to an Ising critical point even after any active processes have likely been disrupted in sample preparation.

Another future direction for this theoretical approach is to convert the phase diagrams using external fields $H$ and $\Delta$ into diagrams based on the concentration of each component. We chose to use a model with fixed external fields and variable composition to enable efficient simulations on small system sizes, and to easily compare with existing theory literature. These external fields could be converted to the corresponding compositions of each component, transforming the phase diagram to one of fixed compositions. This would allow more direct comparison to experimental phase diagrams of model membranes such as in [19].

**Acknowledgements**

EM, DH, and BB were supported by National Institutes of Health grants R01-AI018306 followed by R01-GM117552, and their contributions further benefited from participation in the HHMI/MBL Summer Institute supported by an HCIA award. SW was supported by the Department of Defense through the National Defense Science Engineering Graduate Fellowship (NDSEG) Program. JPS was supported by the National Science Foundation grants DMR-1312160 and DMR-1719490. EM was additionally supported by the National Institutes of Health under the Ruth L. Kirschstein National Research Service Award (2T32GM008267) from the National Institute of General Medical Sciences.

We are grateful to Frank Zhang for discussions on the use of neural networks for phase diagram prediction, to Archishman Raju, Colin Clement, and Benjamin Machta for discussions on critical phenomena and scaling analysis, and to Gerald Feigenson for discussions on lipid membranes.

**References**

(1) Schmid, F. Physical Mechanisms of Micro- and Nanodomain Formation in Multicomponent Lipid Membranes. *Biochim. Biophys. Acta - Biomembr.* **2017**, *1859* (4), 509–528.

(2) Léonard, C.; Alsteens, D.; Dumitru, A. C.; Mingeot-Leclercq, M.-P.; Tyteca, D. Lipid Domains and Membrane (Re)Shaping: From Biophysics to Biology. In *The Biophysics of Cell Membranes: Biological Consequences*; Epand, R. M., Ruysschaert, J.-M., Eds.; Springer Singapore: Singapore, 2017; pp 121–175.

(3) Munro, S. Lipid Rafts. *Cell* **2003**, *115* (4), 377–388.

(4) Pike, L. J. Rafts Defined: A Report on the Keystone Symposium on Lipid Rafts and Cell Function. *J. Lipid Res.* **2006**, *47* (7), 1597–1598.

(5) Sengupta, P.; Holowka, D.; Baird, B. Fluorescence Resonance Energy Transfer between Lipid Probes Detects Nanoscopic Heterogeneity in the Plasma Membrane of Live Cells. *Biophys. J.* **2007**, *92* (10), 3564–3574.

(6) Lingwood, D.; Simons, K. Lipid Rafts as a Membrane-Organizing Principle. *Science* **2010**, *327* (5961), 46–50.

(7) Schick, M. Theories of Equilibrium Inhomogeneous Fluids http://faculty.washington.edu/schick/Abstracts/sens-book.pdf (accessed Jun 6, 2017).

(8) Dietrich, C.; Bagatolli, L. a.; Volovyk, Z. N.; Thompson, N. L.; Levi, M.; Jacobson, K.; Gratton, E. Lipid Rafts Reconstituted in Model Membranes. *Biophys. J.* **2001**, *80* (3), 1417–1428.

(9) Veatch, S. L.; Keller, S. L. Seeing Spots: Complex Phase Behavior in Simple Membranes. *Biochim. Biophys. Acta - Mol. Cell Res.* **2005**, *1746* (3), 172–185.

(10) Veatch, S. L.; Keller, S. L. Separation of Liquid Phases in Giant Vesicles of Ternary Mixtures of Phospholipids and Cholesterol. *Biophys. J.* **2003**, *85* (5), 3074–3083.

(11) Baumgart, T.; Hammond, A. T.; Sengupta, P.; Hess, S. T.; Holowka, D. A.; Baird, B. A.; Webb, W. W. Large-Scale Fluid/fluid Phase Separation of Proteins and Lipids in Giant Plasma Membrane Vesicles. *PNAS* **2007**, *104* (9), 3165–3170.

(12) Machta, B. B.; Papanikolaou, S.; Sethna, J. P.; Veatch, S. L. Minimal Model of Plasma Membrane Heterogeneity Requires Coupling Cortical Actin to Criticality. *Biophys. J.* **2011**, *100* (7), 1668–1677.

(13) Swamy, M. J.; Ciani, L.; Ge, M.; Smith, A. K.; Holowka, D.; Baird, B.; Freed, J. H. Coexisting Domains in the Plasma Membranes of Live Cells Characterized by Spin-Label ESR Spectroscopy. *Biophys. J.* **2006**, *90* (12), 4452–4465.

(14) Simons, K.; Gerl, M. J. Revitalizing Membrane Rafts: New Tools and Insights. *Nat. Rev. Mol. Cell Biol.* **2010**, *11* (10), 688–699.

(15) Holowka, D.; Gosse, J. A.; Hammond, A. T.; Han, X.; Sengupta, P.; Smith, N. L.; Wagenknecht-Wiesner, A.; Wu, M.; Young, R. M.; Baird, B. Lipid Segregation and IgE Receptor Signaling: A Decade of Progress. *Biochim. Biophys. Acta* **2005**, *1746* (3), 252–259.

(16) Sezgin, E.; Levental, I.; Mayor, S.; Eggeling, C. The Mystery of Membrane Organization: Composition, Regulation and Roles of Lipid Rafts. *Nat. Rev. Mol. Cell Biol.* **2017**, *18* (6), 361–374.

(17) Chini, B.; Parenti, M. G-Protein Coupled Receptors in Lipid Rafts and Caveolae: How, When and Why Do They Go There? *J. Mol. Endocrinol.* **2004**, *32* (2), 325–338.

(18) Nussinov, R.; Jang, H.; Tsai, C. J. Oligomerization and Nanocluster Organization Render Specificity. *Biol. Rev.* **2015**, *90* (2), 587–598.

(19) Feigenson, G. W. Phase Diagrams and Lipid Domains in Multicomponent Lipid Bilayer Mixtures. *Biochim. Biophys. Acta* **2009**, *1788* (1), 47–52.

(20) Konyakhina, T. M.; Feigenson, G. W. Phase Diagram of a Polyunsaturated Lipid Mixture: Brain sphingomyelin/1-Stearoyl-2-Docosahexaenoyl-Sn-Glycero-3-Phosphocholine/cholesterol. *Biochim. Biophys. Acta - Biomembr.* **2016**, *1858* (1), 153–161.

(21) Veatch, S. L.; Cicuta, P.; Sengupta, P.; Honerkamp-Smith, A.; Holowka, D.; Baird, B. Critical Fluctuations in Plasma Membrane Vesicles. *ACS Chem. Biol.* **2008**, *3* (5), 287–293.

(22) Stanich, C. A.; Honerkamp-Smith, A. R.; Putzel, G. G.; Warth, C. S.; Lamprecht, A. K.; Mandal, P.; Mann, E.; Hua, T.-A. D.; Keller, S. L. Coarsening Dynamics of Domains in Lipid Membranes. *Biophys. J.* **2013**, *105* (2), 444–454.

(23) Konyakhina, T. M.; Goh, S. L.; Amazon, J.; Heberle, F. A.; Wu, J.; Feigenson, G. W. Control of a Nanoscopic-to-Macroscopic Transition: Modulated Phases in Four-Component DSPC/DOPC/POPC/Chol Giant Unilamellar Vesicles. *Biophys. J.* **2011**, *101* (2), L8–L10.

(24) Levental, I.; Veatch, S. L. The Continuing Mystery of Lipid Rafts. *J. Mol. Biol.* **2016**, *428* (24), 4749–4764.

(25) Schick, M. Membrane Heterogeneity: Manifestation of a Curvature-Induced Microemulsion. *Phys. Rev. E* **2012**, *85* (3), 031902.

(26) Sadeghi, S.; Müller, M.; Vink, R. L. C. Raft Formation in Lipid Bilayers Coupled to Curvature. *Biophys. J.* **2014**, *107* (7), 1591–1600.

(27) Amazon, J. J.; Goh, S. L.; Feigenson, G. W. Competition between Line Tension and Curvature Stabilizes Modulated Phase Patterns on the Surface of Giant Unilamellar

Vesicles: A Simulation Study. *Phys. Rev. E* **2013**, *87* (2), 22708.

(28) Palmieri, B.; Safran, S. A. Hybrid Lipids Increase the Probability of Fluctuating Nanodomains in Mixed Membranes. *Langmuir* **2013**, *29* (17), 5246–5261.

(29) Honerkamp-Smith, A. R.; Veatch, S. L.; Keller, S. L. An Introduction to Critical Points for Biophysicists; Observations of Compositional Heterogeneity in Lipid Membranes. *Biochim. Biophys. Acta* **2009**, *1788* (1), 53–63.

(30) Blank, U.; Rivera, J. The Ins and Outs of IgE-Dependent Mast-Cell Exocytosis. *Trends Immunol.* **2004**, *25* (5), 266–273.

(31) Gilfillan, A. M.; Rivera, J. The Tyrosine Kinase Network Regulating Mast Cell Activation. *Immunol. Rev.* **2009**, *228* (1), 149–169.

(32) Rivera, J.; Gilfillan, A. M. Molecular Regulation of Mast Cell Activation. *J. Allergy Clin. Immunol.* **2006**, *117* (6), 1214–1225.

(33) Holowka, D.; Baird, B. Roles for Lipid Heterogeneity in Immunoreceptor Signaling. *Biochim. Biophys. Acta - Mol. Cell Biol. Lipids* **2016**, *1861* (8, Part B), 830–836.

(34) Shelby, S. A.; Veatch, S. L.; Holowka, D. A.; Baird, B. A. Functional Nanoscale Coupling of Lyn Kinase with IgE-FcεRI Is Restricted by the Actin Cytoskeleton in Early Antigen-Stimulated Signaling. *Mol. Biol. Cell* **2016**, *27* (22), 3645–3658.

(35) Gompper, G.; Schick, M. Lattice Model of Microemulsions: The Effect of Fluctuations in One and Two Dimensions. *Phys. Rev. A* **1990**, *42* (4), 2137–2149.

(36) Gompper, G.; Schick, M. Lattice Model of Microemulsions. *Phys. Rev. B* **1990**, *41* (13), 9148–9162.

(37) Blume, M.; Emery, V. J.; Griffiths, R. B. Ising Model for the Λ Transition and Phase Separation in He3-He4 Mixtures. *Phys. Rev. A* **1971**, *4* (3), 1071–1077.

(38) Blume, M. Theory of the First-Order Magnetic Phase Change in UO2. *Phys. Rev.* **1966**, *141* (2), 517–524.

(39) Capel, H. W. On the Possibility of First-Order Transitions in Ising Systems of Triplet Ions with Zero-Field Splitting. *Physica* **1966**, *32* (5), 966–988.

(40) Krizhevsky, A.; Sutskever, I.; Hinton, G. E. *ImageNet Classification with Deep Convolutional Neural Networks*; Pereira, F., Burges, C. J. C., Bottou, L., Weinberger, K. Q., Eds.; Curran Associates, Inc., 2012.

(41) Carrasquilla, J.; Melko, R. G. Machine Learning Phases of Matter. *Nat. Phys.* **2017**, *13* (5), 431–434.

(42) Wang, L. Discovering Phase Transitions with Unsupervised Learning. *Phys. Rev. B - Condens. Matter Mater. Phys.* **2016**, *94* (19), 195105.

(43) Gompper, G.; Schick, M. Lattice Model of Microemultions. *Phys. Rev. B* **1990**, *41* (13), 9148–9162.

(44) Machta, B. B.; Veatch, S. L.; Sethna, J. P. Critical Casimir Forces in Cellular Membranes. *Phys. Rev. Lett.* **2012**, *109* (13), 1–5.

(45) Nielsen, M. A. Neural Networks and Deep Learning neuralnetworksanddeeplearning.com (accessed Jan 1, 2017).

(46) Bennett, C. H. Efficient Estimation of Free Energy Differences from Monte Carlo Data. *J. Comput. Phys.* **1976**, *22*, 245–268.

(47) Metropolis, N.; Rosenbluth, A. W.; Rosenbluth, M. N.; Teller, A. H.; Teller, E. Equation of State Calculations by Fast Computing Machines. *J. Chem. Phys.* **1953**, *21* (6), 1087–1092.

(48) Jarzynski, C. Nonequilibrium Equality for Free Energy Differences. *Phys. Rev. Lett.* **1997**, *78* (14), 2690–2693.

(49) Beale, P. D. Finite-Size Scaling Study of the Two-Dimensional Blume-Capel Model. *Phys. Rev. B* **1986**, *33* (3), 1717–1720.

(50) Cardy, J. *Scaling and Renormalization in Statistical Physics*; Cambridge University Press: New York, New York, USA, 1996.

(51) Palmieri, B.; Grant, M.; Safran, S. A. Prediction of the Dependence of the Line Tension on the Composition of Linactants and the Temperature in Phase Separated Membranes. *Langmuir* **2014**, *30* (39), 11734–11745.

(52) Sil, D.; Lee, J. B.; Luo, D.; Holowka, D.; Baird, B. Trivalent Ligands with Rigid DNA Spacers Reveal Structural Requirements for IgE Receptor Signaling in RBL Mast Cells. *ACS Chem. Biol.* **2007**, *2* (10), 674–684.

(53) Kusumi, A.; Fujiwara, T. K.; Morone, N.; Yoshida, K. J.; Chadda, R.; Xie, M.; Kasai, R. S.; Suzuki, K. G. N. Membrane Mechanisms for Signal Transduction: The Coupling of the Meso-Scale Raft Domains to Membrane-Skeleton-Induced Compartments and Dynamic Protein Complexes. *Semin. Cell Dev. Biol.* **2012**, *23* (2), 126–144.

(54) Honigmann, A.; Sadeghi, S.; Keller, J.; Hell, S. W.; Eggeling, C.; Vink, R. A Lipid Bound Actin Meshwork Organizes Liquid Phase Separation in Model Membranes. *eLife* **2014**, *3*, e01671.

(55) Nienhuis, B.; Berker, A. N.; Riedel, E. K.; Schick, M. First- and Second-Order Phase Transitions in Potts Models: Renormalization-Group Solution. *Phys. Rev. Lett.* **1979**, *43* (11), 737–740.

(56) Pearson, R. B. Conjecture for the Extended Potts Model Magnetic Eigenvalue. *Phys. Rev. B* **1980**, *22* (5), 2579–2580.

(57) Nienhuis, B.; Warnaar, S. O.; Blote, H. W. J. Exact Multicritical Behaviour of the Potts Model. *J. Phys. A. Math. Gen.* **1993**, *26* (3), 477.

(58) Kwak, W.; Jeong, J.; Lee, J.; Kim, D.-H. First-Order Phase Transition and Tricritical Scaling Behavior of the Blume-Capel Model: A Wang-Landau Sampling Approach. *Phys. Rev. E. Stat. Nonlin. Soft Matter Phys.* **2015**, *92* (2-1), 22134.

(59) Sherman, E.; Barr, V.; Samelson, L. E. Super-Resolution Characterization of TCR-Dependent Signaling Clusters. *Immunol. Rev.* **2013**, *251* (1), 21–35.

(60) Rao, M.; Mayor, S. Active Organization of Membrane Constituents in Living Cells. *Curr. Opin. Cell Biol.* **2014**, *29*, 126–132.

(61) Noble, A. E.; Machta, J.; Hastings, A. Emergent Long-Range Synchronization of Oscillating Ecological Populations without External Forcing Described by Ising Universality. *Nat. Commun.* **2015**, *6*, 6664.

# Supporting Information

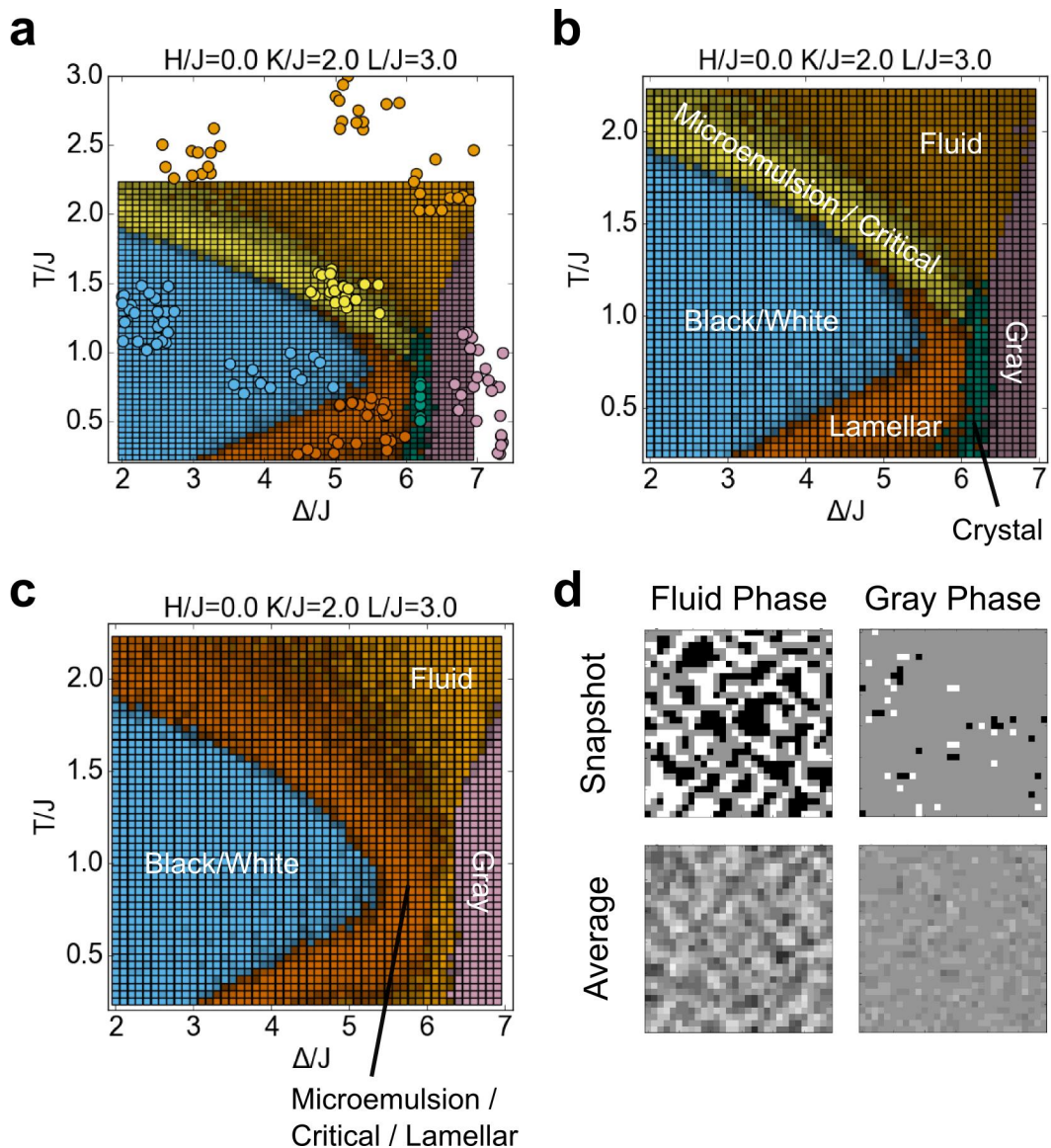

**Figure S1:** Training the neural network. (a) Each circle at specified (*Δ/J*, *T/J*) coordinates represents a parameter set at which training data were acquired. The color indicates the human-generated classification of the phase at that point. These points are overlaid on the final phase diagram computed by the neural network. (b-c) Phase diagrams generated by two types of neural network. (b) is trained on averages of correlated Monte Carlo snapshots to identify all six phases, while (c) is trained on individual Monte Carlo snapshots to identify four

phases. Our final phase diagram is based mostly on (b), but we take the classification of the gray phase from (c) due to its high confidence in that phase. (d) Example simulation images taken near the Fluid / Gray boundary (Fluid at $T/J$ = 1.49 , $\Delta/J$ = 6.2 ; Gray at $T/J$ = 1.49 , $\Delta/J$ = 6.8), demonstrating the one case in which working with snapshots (phase diagram (c)) is advantageous compared to working with averages (phase diagram (d)). Note that the snapshots shown are easily distinguishable as two different phases, but the averages shown look more similar to each other.

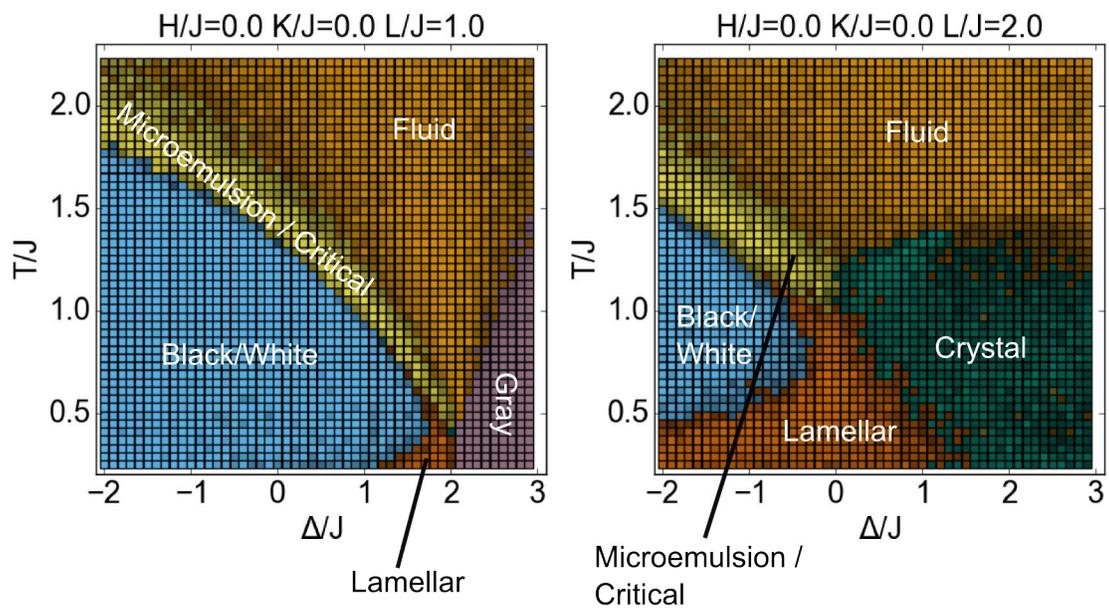

**Figure S2:** Cross-sections of the phase diagram in which H/J=0, K/J=0, and L/J = 1 or 2. Colors have the same meaning as in Figure 3 and Figure S1.

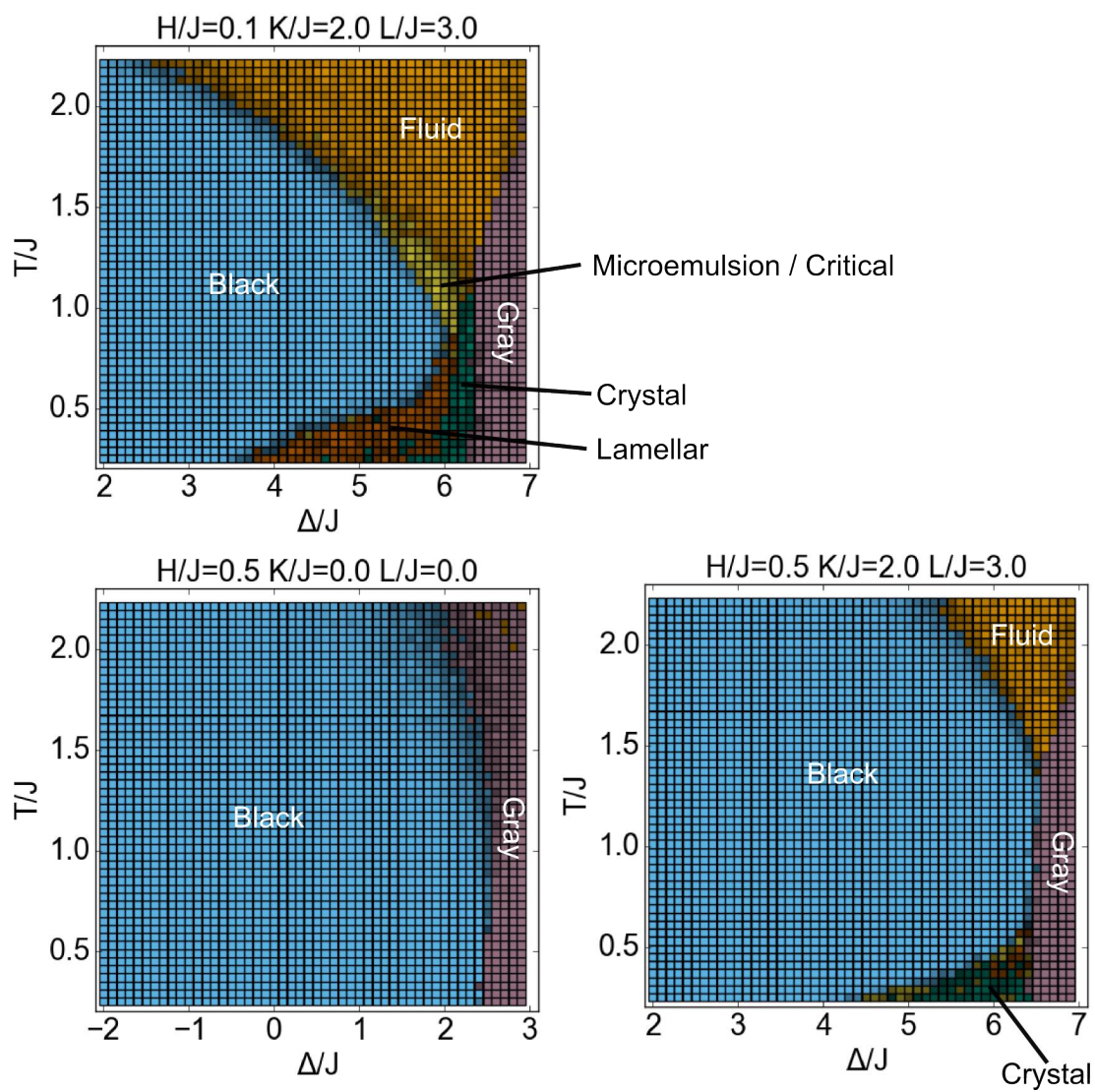

**Figure S3:** Additional cross-sections of the phase diagram at positive H. Colors have the same meaning as in Figure 3 and Figure S1.

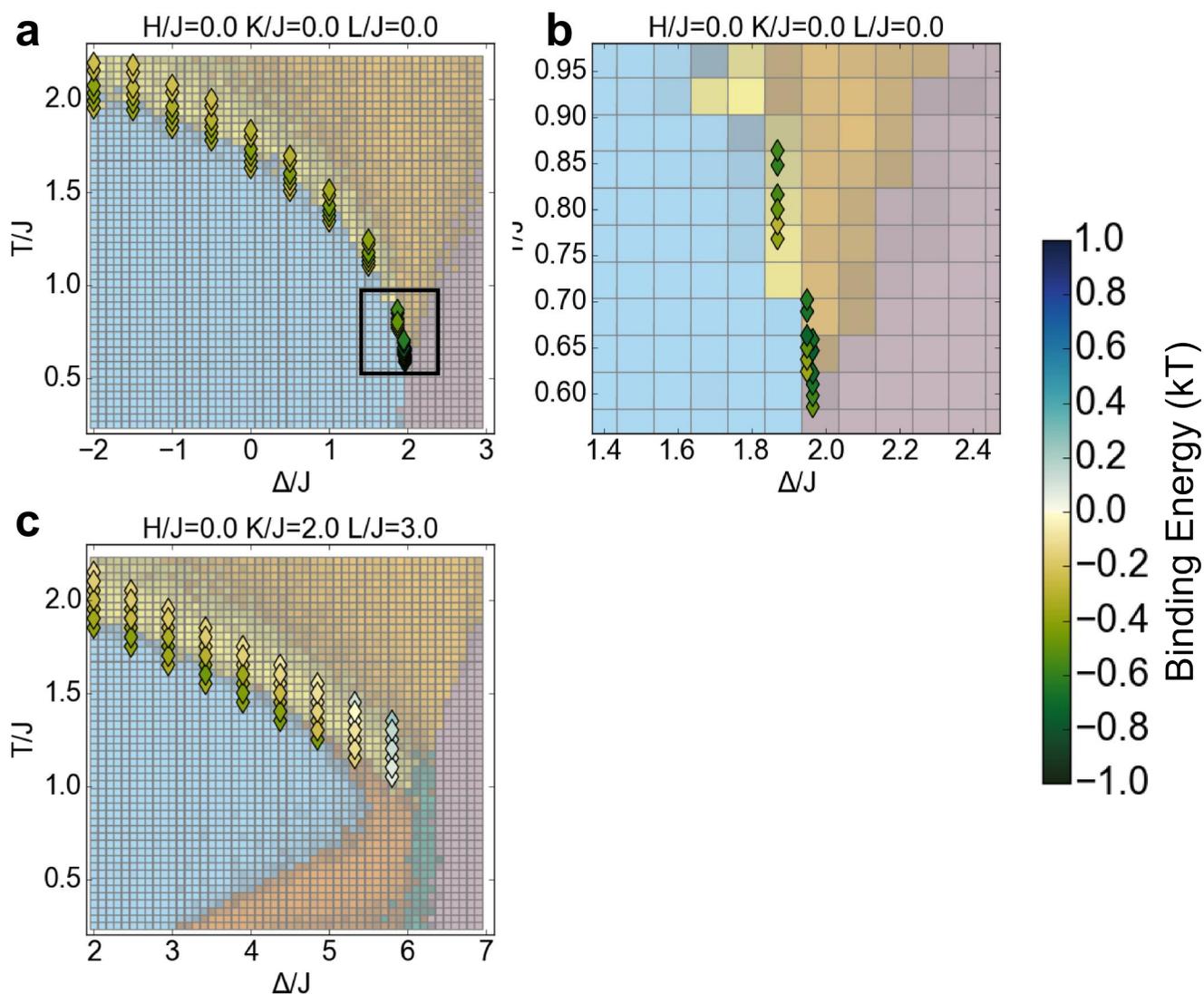

**Figure S4:** Free energy change (binding energy) due to moving two white pixels to a spacing of 3 lattice units apart. Simulations were run on a system similar to that shown in Figure 5, but with the box on the left containing, instead of the set cluster of 3 IgE receptors (each represented by 12 pixels), a single fixed white pixel. Each diamond indicates the free energy calculated by Bennett's method at that point in the phase diagram. (a) Binding energy in the Blume-Capel model (*H*=0, *K*=0, *L*=0) cross-section. (b) Inset of (a) in the area of the tricritical point (black box in (a)). Similar to the more complicated case of clustered receptors (Figure 6a-b), the binding energy is considerably higher close to the tricritical point. (c) Binding energy

in a cross-section that includes microemulsions ($K/J$ = 2, $L/J$ = 3). Similar to the more complicated case (Figure 6d), certain microemulsions give a positive binding energy. We provide data for this case, which is simpler than that shown in the main text, in the hopes that it will prove useful for future theoretical work related to this model. In particular, we note that it may be possible develop a universal scaling theory to describe the increase in binding energy magnitude as the tricritical point is approached.

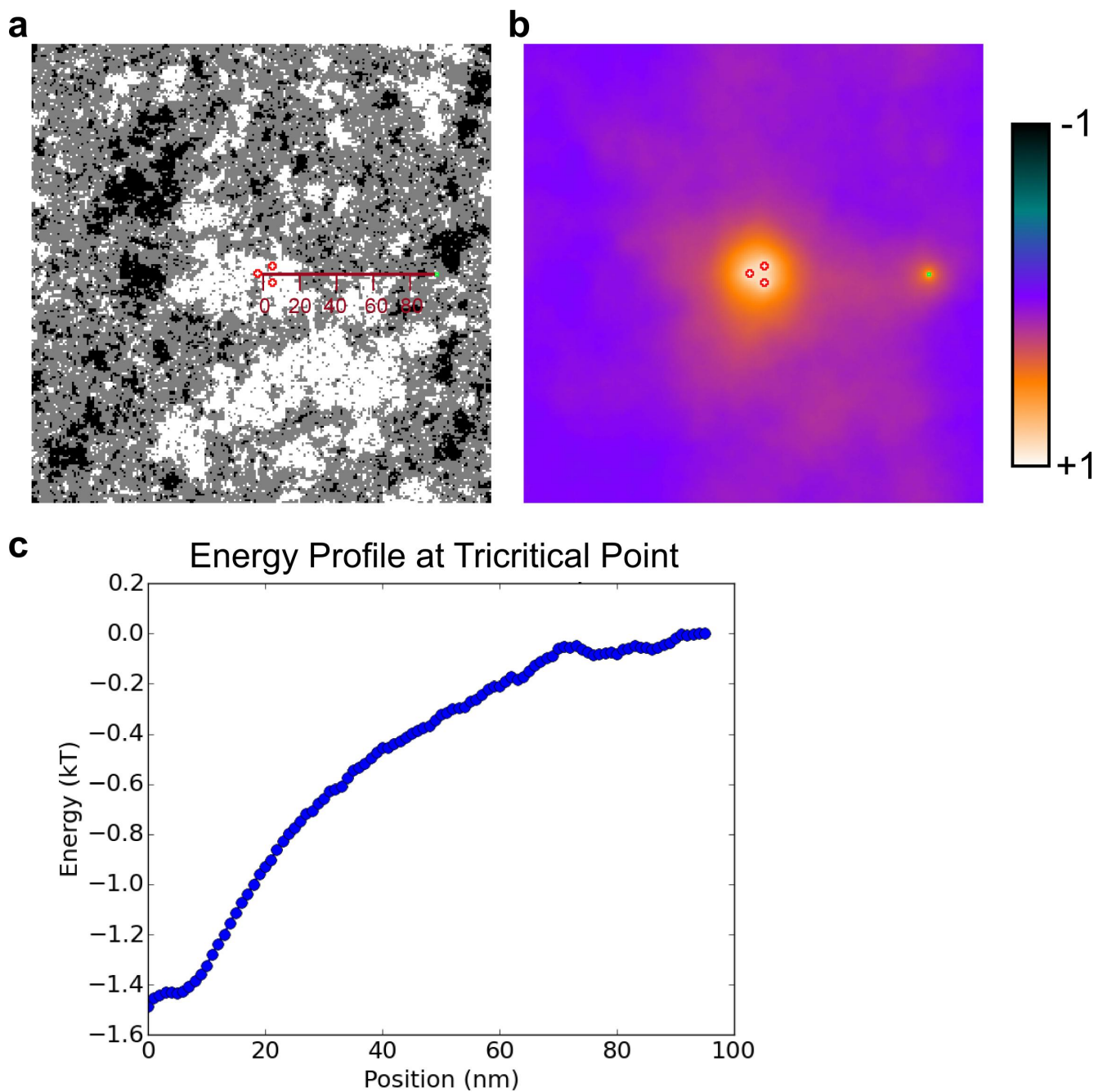

**Figure S5:** Alternate method for calculating kinase binding energy. We use the method described in [1]. (a-b) Example simulation at the tricritical point, in which the kinase sits at a position 95 nm from the cluster. As in Figure 5, receptors (red) and kinase (green) were held fixed as white, and the rest of the lattice was Monte Carlo sampled. (a) shows one simulation snapshot, and (b) shows the average of all snapshots acquired. The simulation was repeated

at every possible kinase position from 0 to 95, stepwise, in order to generate (c), the profile showing the free energy of the kinase in every position. The position axis in (c) corresponds to the axis shown in (a). The free energy at position 3 nm of ~ -1.4 $k_BT$ is in good agreement with the computationally cheaper method used in the rest of the study (Figure 6, Figure S4).

a

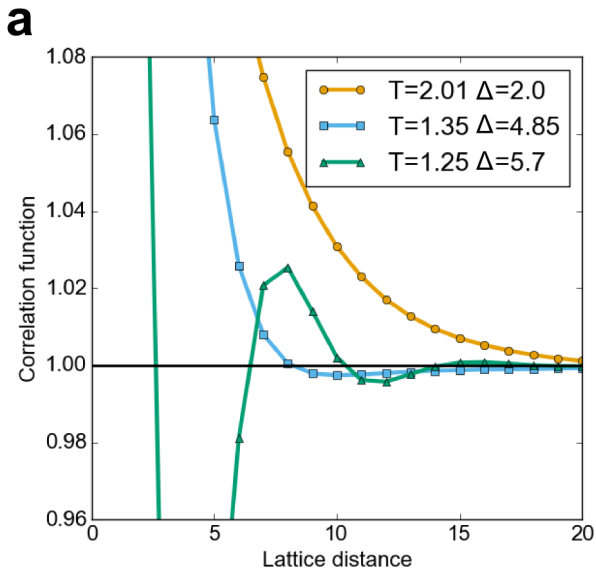

b

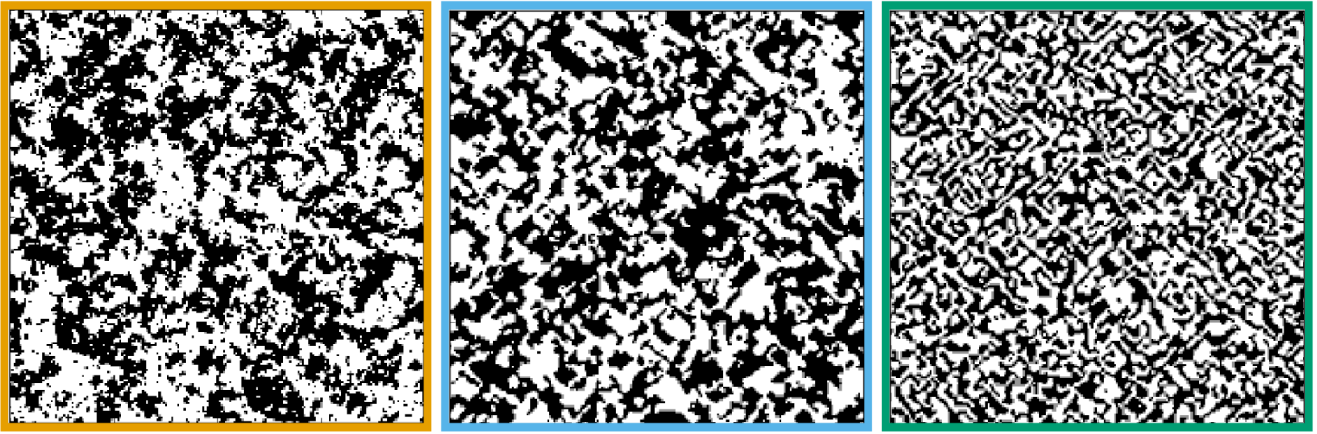

**Figure S6:** Detailed analysis of three selected points in the *H/J*=0, *K/J*=2, *L/J*=3 cross-section. (a) Correlation functions between white pixels at the selected points. The 2D correlation functions *G* are calculated according to the formula

$$G = \frac{1}{\rho^2} \frac{FFT^{-1}(|FFT(A)|^2)}{FFT^{-1}(|FFT(M)|^2)}$$

where *A* is a 2D array with value 1 if the σ value at the corresponding pixel of the snapshot is 1 (white pixel), or 0 otherwise, *M* is an array of all 1's of the same size as *A*, *ρ* is the density of white pixels (calculated as the sum of $A$ divided by the sum of $M$), and FFT represents the fast Fourier transform. A correlation function value greater than 1 indicates positive correlation, 1 indicates no correlation, and less than 1 indicates anticorrelation. The function plotted here is the average transect of $G$ along the horizontal (1,0) and vertical (0,1) directions (used instead of a radial average because $G$ is not always radially symmetric). Correlation functions are averaged over 2000 snapshots of size 200x200 per point.

One criterion to define a microemulsion is that this correlation function must have a local minimum, rather than decrease monotonically. By this criterion, $T=2.01$, $\Delta=2.0$ is not a microemulsion, as the correlation function decreases monotonically. $T=1.35$, $\Delta=4.85$ is a microemulsion with length scale ~ 10 because the first local minimum occurs at lattice distance 10, and $T=1.25$, $\Delta=5.7$ is a microemulsion with length scale ~ 4 because the (considerably stronger) first local minimum occurs at lattice distance 4. (b) The analysis using correlation functions gives results consistent with a visual inspection of snapshots at the same three points, which appear as critical fluctuations (left), a long length-scale microemulsion (center), and a short length-scale microemulsion (right).

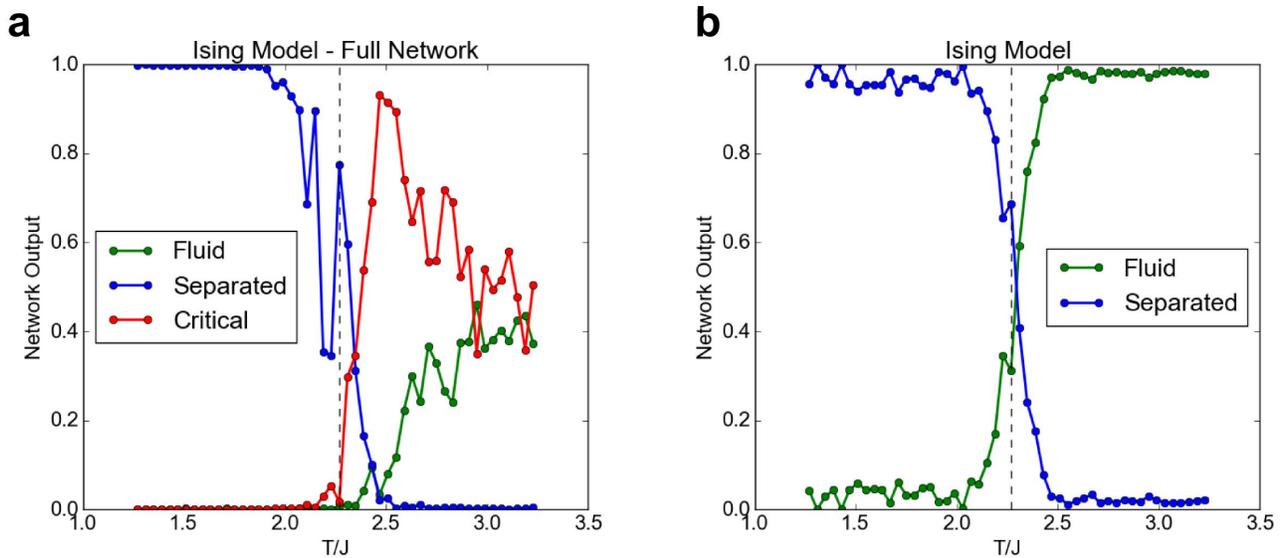

**Figure S7:** Neural network computation of the Ising model phase diagram. Graphs show the outputs of the neural network corresponding to each phase, indicating the network's level of confidence that the system represents that phase. At each temperature value, the network's prediction is the phase for which the network output is the highest. The true Ising transition temperature is indicated by the dashed line; an accurate network should give "Separated" as the highest output below this temperature, and "Critical" or "Fluid" as the highest output above this temperature. (a) Output of the neural network used in the rest of this study. The predicted transition temperature is close to, but slightly above, the true transition temperature. (b) Output of a neural network specialized for the Ising model only. This network was trained on examples from the Ising model, and only distinguishes between the fluid and phase-separated states. This network's predicted transition temperature almost exactly matches the true transition temperature.

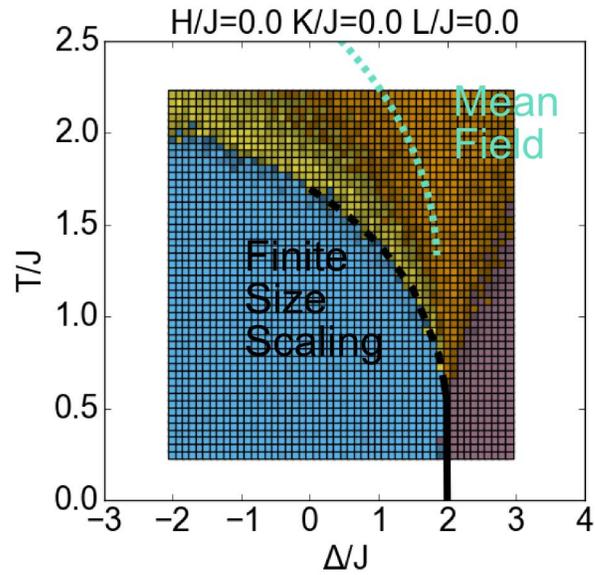

**Figure S8:** Comparison to other methods for calculation of the Blume-Capel phase diagram. The neural network's phase diagram follows the same key as in Figure 2. Overlaid in gray is the phase boundary as determined by finite size scaling in [2]. The dashed line represents a critical phase transition, solid represents a first-order transition, and the transition from dashed to solid is the tricritical point. Dotted line in blue is the critical line as determined by mean field theory in [3].

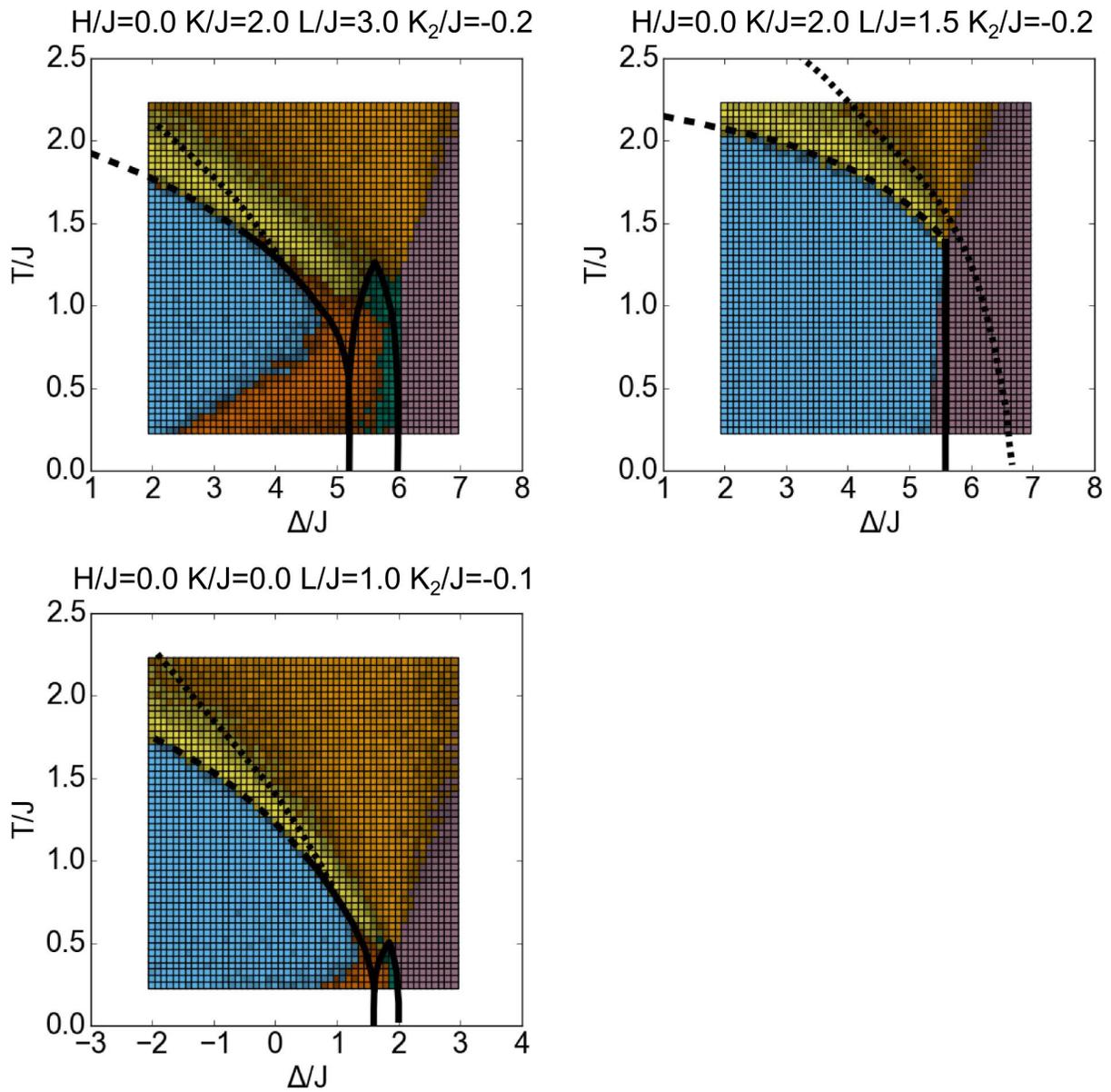

**Figure S9:** Comparing phase diagram results from our study to the original study. Gray overlays are reproduced from the original study.[4] Solid lines indicate first order phase transitions, dashed lines indicate critical phase transitions, and dotted lines are Lifshitz lines.

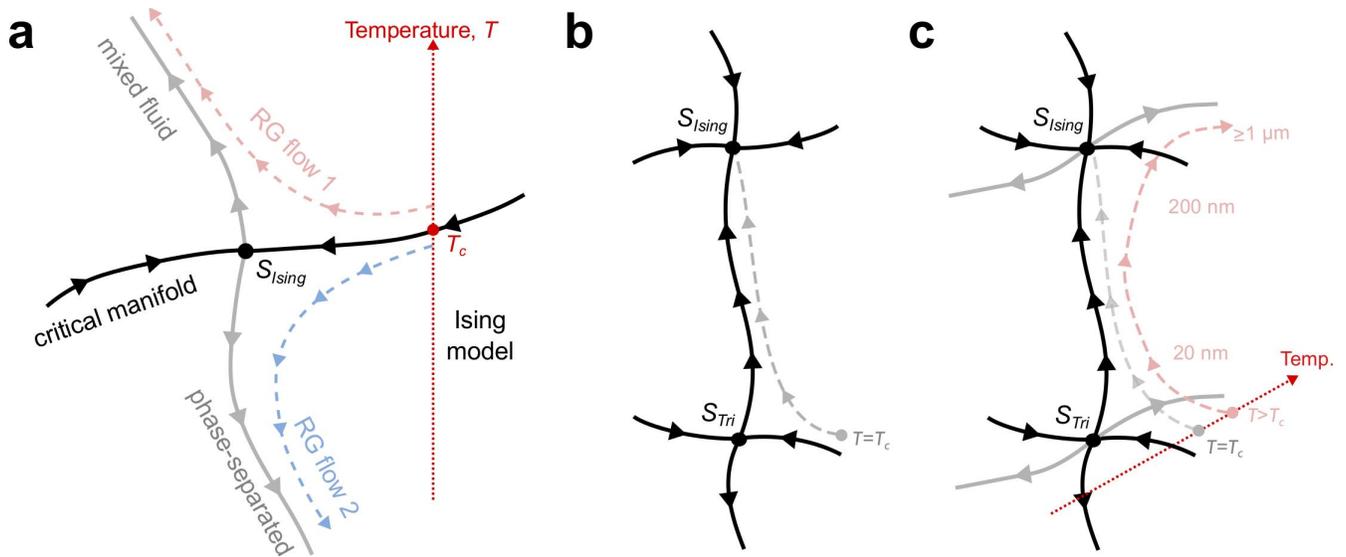

**Figure S10:** Schematic of renormalization group flows and crossover behavior. The operation of coarse-graining leads to flows through the space of possible models (i.e. a space whose different directions correspond to different model parameters), a process formalized by the renormalization group (RG). Fixed points in this model space are models that look the same at all length scales, and correspond to different classes of critical phenomena. (a) gives an example of RG flows in the familiar Ising model. $S_{Ising}$ is the Ising critical fixed point; under coarse-graining, models flow towards it along black lines and away along gray lines. For the Ising model, temperature (vertical, dotted red arrow) is crucial in determining model behavior. Above the critical temperature, $T_c$, as we coarse-grain our model (RG Flow 1), it flows to a mixed (uniform) fluid state. Below $T_c$ (RG Flow 2), the model flows toward a phase-separated state. Importantly, when the models in either RG Flow 1 or 2 pass near $S_{Ising}$, they exhibit Ising-like properties, and coarse-graining to length scales of 20 nm is typically sufficient to produce critical behavior. (b) In the presence of more than one critical fixed point, a model's trajectory through RG flows can pass near—and thus be influenced by—the multiple different fixed points. In this schematic, we imagine the trajectory of a membrane model in such a space (gray dashed arrow), where $S_{Tri}$ and $S_{Ising}$ are the 2D tricritical and 2D Ising critical fixed

S14

points (respectively). At $T=T_c$, where $T_c$ is the Ising critical temperature, the model flows directly to the Ising fixed point. The transition from tricritical-like to Ising-like behavior is referred to as crossover. (c) In the same space as (b), above $T_c$, our model can flow along a complicated trajectory (red dashed arrow line). Note that here we have included a 3$^{rd}$ dimension—temperature, which acts as in (a). At short length scales (~20 nm), our model is near the tricritical fixed point, and will thus obey tricritical behavior. As we coarse-grain further, our model flows away toward the Ising critical point, so that the behavior of the model at intermediate length scales (~200 nm) is Ising-like. As we coarse-grain even further, the model flows away to some non-critical fixed point.

**Supporting References**

(1) Machta, B. B.; Veatch, S. L.; Sethna, J. P. Critical Casimir Forces in Cellular Membranes. *Phys. Rev. Lett.* **2012**, *109* (13), 138101.

(2) Beale, P. D. Finite-Size Scaling Study of the Two-Dimensional Blume-Capel Model. *Phys. Rev. B* **1986**, *33* (3), 1717–1720.

(3) Blume, M.; Emery, V. J.; Griffiths, R. B. Ising Model for the λ Transition and Phase Separation in He3-He4 Mixtures. *Phys. Rev. A* **1971**, *4* (3), 1071–1077.

(4) Gompper, G.; Schick, M. Lattice Model of Microemulsions: The Effect of Fluctuations in One and Two Dimensions. *Phys. Rev. A* **1990**, *42* (4), 2137–2149.